\documentclass[twoside,twocolumn,reprint,secnumarabic,superscriptaddress,amssymb,amsmath,nobibnotes,nofootinbib,aps,floatfix]{revtex4-1}
\usepackage[latin9]{inputenc}
\usepackage{color}
\usepackage{amsmath}
\usepackage{graphicx}
\usepackage{wasysym}
\usepackage[unicode=true,pdfusetitle,
 bookmarks=true,bookmarksnumbered=false,bookmarksopen=false,
 breaklinks=true,pdfborder={0 0 0},pdfborderstyle={},backref=false,colorlinks=true]
 {hyperref}
\hypersetup{
 linkcolor=blue,citecolor=blue,urlcolor=blue}

\makeatletter
\graphicspath{{figures/}}
\usepackage{float}
\usepackage{tikz}
\usetikzlibrary{calc,shapes,backgrounds}
\usetikzlibrary{datavisualization,patterns}
\usetikzlibrary{arrows.meta}
\usepackage{pslatex}
\usepackage{color}

\makeatother

\begin{document}
\title{The correlation functions of certain random antiferromagnetic spin-$1/2$
critical chains}
\author{João C. Getelina}
\affiliation{Instituto de F\'{i}sica de São Carlos, Universidade de São Paulo,
CP 369, 13560-970, São Carlos, SP, Brazil}
\affiliation{Department of Physics, Missouri University of Science and Technology,
Rolla, MO 65409, USA}
\author{José A. Hoyos}
\affiliation{Instituto de F\'{i}sica de São Carlos, Universidade de São Paulo,
CP 369, 13560-970, São Carlos, SP, Brazil}
\date{\today}
\begin{abstract}
We study the spin-spin correlations in two distinct random critical
XX spin-1/2 chain models via exact diagonalization. For the well-known
case of uncorrelated random coupling constants, we study the non-universal
numerical prefactors and relate them to the corresponding Lyapunov
exponent of the underlying single-parameter scaling theory. We have
also obtained the functional form of the correct scaling variables
important for describing even the strongest finite-size effects. Finally,
with respect to the distribution of the correlations, we have numerically
determined that they converge to a universal (disorder-independent)
non-trivial and narrow distribution when properly rescaled by the
spin-spin separation distance in units of the Lyapunov exponent. With
respect to the less known case of correlated coupling constants, we
have determined the corresponding exponents and shown that both typical
and mean correlations functions decay algebraically with the distance.
While the exponents of the transverse typical and mean correlations
are nearly equal, implying a narrow distribution of transverse correlations,
the longitudinal typical and mean correlations critical exponents
are quite distinct implying much broader distributions. Further comparisons
between these models are given.
\end{abstract}
\maketitle

\section{Introduction\label{sec:intro} }

Random quantum spin chains have proved to be a fruitful platform for
developing new methodologies and fundamental concepts in condensed
matter physics. One of the most successful methods developed so far
is the so-called strong-disorder renormalization-group (SDRG) method~\cite{Ma1979,Dasgupta1980,bhatt-lee},
which has been applied to a plethora of random systems (see Refs.~\onlinecite{Igloi2005,Igloi2018}
for reviews). Inherently linked to it is the concept of infinite-randomness
fixed points~\cite{Fisher1992,Fisher1994}. These are critical points
in which the statistical fluctuations of local quantities, surprisingly,
increase without limits along the renormalization-group flow yielding
to an exotic type of activated dynamical scaling. Equally important,
due to the unbounded increase of the statistical fluctuations, the
SDRG method is believed to exactly capture the universal properties
of these fixed points. Finally, it is noteworthy that these fixed
points control the phase transitions and critical phases of many quantum,
classical and non-equilibrium disordered systems (see Ref.~\onlinecite{vojta-review06}
for a review).

In this context, the random antiferromagnetic quantum spin-1/2 chain
is a paradigmatic model which for many years has been stimulating
theoretical~\cite{Ma1979,Dasgupta1980,Doty1992,Fisher1994,Laflorencie2004,hoyos-rigolin,Hoyos2007,Giamarchi2012,Shu2016,Getelina2018,Sandvik2018}
and experimental~\cite{theodorou,Masuda2004,Mesot2013,Mesot2018}
studies. For a large range of anisotropies, it is a critical system
governed by an infinite-randomness fixed point amenable to many analytical
predictions of the SDRG method. A striking one is that the average
value spin-spin correlations decays algebraically with the distance
$\sim r^{-\eta_{\alpha}}$ with universal (disorder-independent) isotropic
exponent $\eta_{x}=\eta_{z}=2$, while the typical value decays stretched
exponentially fast $\sim e^{-\sqrt{r}}$~\cite{Fisher1994}.

Nonetheless, this knowledge is far from satisfactory when compared
to the clean chain. Not only the exponents of the leading and subleading
terms are know, but also the corresponding numerical prefactors~\cite{Lieb1961,mccoy-68,luther-peschel-xxz,haldane-prl80,Giamarchi1989,Singh1989,Hallberg1995,lukyanov-zamolodchikov-97,Affleck1998,Lukyanov1998,Hikihara1998,lukyanov-prefactor}.
It is the purpose of this work to shorten the knowledge gap between
clean and disordered systems by studying non-universal (disorder-dependent)
details of the spin-spin correlation functions, such as the numerical
prefactors and scaling variables.

Recently, it was discovered that the paradigmatic random antiferromagnetic
quantum spin-1/2 chain can also be governed by a line of finite-disorder
fixed points when a certain type of correlations are present in the
random coupling constants~\cite{Hoyos2011,Getelina2016,Getelina2018}.
This is an exciting result not only because it allows us studying
new physical phenomena in a simple and well-known model, but also
because the correlations among the disorder variables are the same
present in a class of polymers~\cite{Epstein1987,Dunlap1990,Phillips1991}.
However, unlike the uncorrelated disorder model, much less is known
about its average correlation function. Nothing about the typical
correlations are known. For this reason, it is also the purpose of
this work to study the corresponding critical exponents.

In Sec.~\ref{sec:methods}, we define the models studied, review
further relevant results for our purposes, and provide the methodology
of our study. In Secs.~\ref{sec:uncorr} and \ref{sec:corr} we report
our results on the correlation functions of the uncorrelated and correlated
disordered spin chains, respectively. Finally, we provide further
discussions and concluding remarks to Sec.~\ref{sec:conclusion}.

\section{Models, known results and methods\label{sec:methods}}

In this section we define the studied models, review key known results
in the literature about the spin-spin correlation functions, and explain
our methods.

\subsection{Models }

The Hamiltonian of the random XXZ spin-$1/2$ chain is 
\begin{equation}
H=\sum_{i=1}^{L}J_{i}\left(S_{i}^{x}S_{i+1}^{x}+S_{i}^{y}S_{i+1}^{y}+\Delta S_{i}^{z}S_{i+1}^{z}\right),\label{eq:H}
\end{equation}
where $S_{i}^{\alpha}$ are spin-$1/2$ operators, $J_{i}$ are the
random coupling constants, and $\Delta$ is the anisotropy parameter.
We consider chains of even size $L$ with periodic boundary conditions
$S_{i+L}^{\alpha}=S_{i}^{\alpha}$. The coupling constants $J_{i}$
are realizations of a random variable drawn from the probability distribution
\begin{equation}
P(J)=\begin{cases}
\frac{1}{D}J^{\frac{1}{D}-1}, & \text{if}\;0<J<1\\
0, & \text{otherwise}.
\end{cases}\label{eq:P-J}
\end{equation}
 Here, the disorder strength is parameterized by $D\geq0$, with $D=0$
representing the uniform (clean) system and $D\rightarrow\infty$
representing an infinitely disordered system. In addition, we consider
the cases of (i) uncorrelated couplings $\overline{J_{i}J_{k}}=\overline{J_{i}}\times\overline{J_{k}}$
and (ii) perfectly and locally correlated couplings such that the
coupling sequence is $\{J_{1}J_{1}J_{2}J_{2}\dots J_{\frac{L}{2}}J_{\frac{L}{2}}\}$,
with $\overline{J_{i}J_{k}}=\overline{J_{i}}\times\overline{J_{k}}$.

Finally, in this work we will consider only the $\Delta=0$ case.

\subsection{Some known results for the case of uncorrelated couplings}

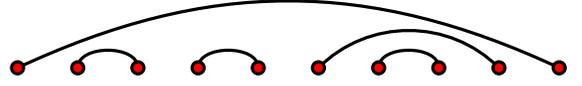
\begin{figure}[t]
\begin{centering}
\begin{tikzpicture}[scale=0.8,clip]
\draw [very thick] (12,6) to [out=25,in=155] (21,6);
\draw [very thick] (13,6) to [out=90,in=90] (14,6);
\draw [very thick] (15,6) to [out=90,in=90] (16,6);
\draw [very thick] (17,6) to [out=45,in=135] (20,6);
\draw [very thick] (18,6) to [out=90,in=90] (19,6);
\draw [fill=red,very thick] (12,6) circle [radius=.1];
\draw [fill=red,very thick] (13,6) circle [radius=.1];
\draw [fill=red,very thick] (14,6) circle [radius=.1];
\draw [fill=red,very thick] (15,6) circle [radius=.1];
\draw [fill=red,very thick] (16,6) circle [radius=.1];
\draw [fill=red,very thick] (17,6) circle [radius=.1];
\draw [fill=red,very thick] (18,6) circle [radius=.1];
\draw [fill=red,very thick] (19,6) circle [radius=.1];
\draw [fill=red,very thick] (20,6) circle [radius=.1];
\draw [fill=red,very thick] (21,6) circle [radius=.1];
	\end{tikzpicture} 
\par\end{centering}
\caption{\label{fig:RS-state}Schematic of the random-singlet state, which
gives the approximate ground state of the Hamiltonian \eqref{eq:H}
for $-\frac{1}{2}<\Delta\protect\leq1$, according to the strong-disorder
renormalization group method.}
\end{figure}

For uncorrelated random couplings, the SDRG method predicts that the
low-energy critical physics of \eqref{eq:H} is governed by an infinite-randomness
critical fixed point for $-\frac{1}{2}<\Delta\le1$~\cite{Doty1992,Fisher1994}.
It is universal in the sense that the corresponding singular behavior
does not depend on $P(J)$ provided that $P(J<0)=0$ and it is not
excessively singular at $J=0$~\cite{Fisher1994}. In addition, the
method predicts that a good approximation of the corresponding ground
state is the random-singlet state (as depicted in Fig.~\ref{fig:RS-state})
from which much information about the physics can be obtained.

The first one is that spin pairs become locked into SU(2)-symmetric
singlet states, and thus, the bare SO(2) symmetry of \eqref{eq:H}
is enhanced to SU(2). As a consequence, the universal properties of
the system become SU(2) isotropic.~\footnote{This phenomenon of symmetry enhancing is known to be general in random
antiferromagnetic SO($N$) spin chains exhibiting SU($N$) symmetric
singular properties~\cite{quito-hoyos-miranda-prl15,quito-etal-epjb20,quito-etal-prb19}.}

Another useful information is related to the distribution of the singlet
lengths which decays as $\frac{2}{3}r^{-2}$~\cite{Hoyos2007} for
lengths $1\ll r\ll L$. Since those singlets are strongly correlated,
they dominate the (arithmetic average) mean spin-spin correlation
function. Thus, 
\begin{equation}
\overline{C^{\alpha\alpha}}(r)\equiv\overline{\left\langle S_{i}^{\alpha}S_{i+r}^{\alpha}\right\rangle }=\frac{(-1)^{r}}{12r^{\eta}}\times\begin{cases}
c_{\text{o},\alpha}, & \text{for \ensuremath{r} odd}\\
c_{\text{e},\alpha}, & \text{for \ensuremath{r} even}
\end{cases},\label{eq:mean-C}
\end{equation}
 with universal and isotropic exponent $\eta=2$, and non-universal
and anisotropic multiplicative constants $c_{\text{o,e},\alpha}\geq0$.~\footnote{For $\Delta=0$, the longitudinal correlation between spins in the
same sublattice vanishes, and thus, $c_{\text{e},z}=0$~\cite{Lieb1961}.} Surprisingly, it was conjectured~\cite{Hoyos2007} that $c_{\text{o},\alpha}-c_{\text{e},\alpha}=1$
is universal for $\alpha$ being a symmetry axis, i.e., for $\alpha=z$,
and for any $\alpha$ when $\Delta=1$. (Here, $\left\langle \cdots\right\rangle $
and $\overline{\cdots}$ denote the quantum and the disorder averages,
respectively.)

The universality of the exponent $\eta$ was disputed some years ago~\cite{hamacher-etal-prl02},
but there is now a consensus that this is an exact result~\cite{Henelius1998,Igloi2000,Laflorencie2004,xavier-etal-prb18}.
Evidently, numerical confirmations of the constants $c_{\text{e,o},\alpha}$
are much more difficult~\cite{Hoyos2007,Getelina2016}.

We note that logarithmic corrections to \eqref{eq:mean-C} have been
reported in numerical studies for the free-fermion case $\Delta=0$~\cite{Igloi2000}
in which 
\begin{equation}
\overline{C^{xx}}\sim\left(r^{\eta}\ln r\right)^{-1},\label{eq:mean-Cx-Igloi}
\end{equation}
 and for the Heisenberg case $\Delta=1$~\cite{Sandvik2018} in which~\footnote{While the SDRG method have been believed to deliver asymptotic exact
results for the XXZ spin chain \eqref{eq:H}, these numerical results
cast some doubts on this belief. As shown latter, our results do not
exhibit any logarithmic correction.} 
\begin{equation}
\overline{C^{\alpha\alpha}}\sim r^{-\eta}\sqrt{\ln r/r_{0}}.\label{eq:mean-C-Sandvik}
\end{equation}

In contrast, the (geometric average) typical spin-spin correlation
function behaves completely different since the spin pairs are weakly
coupled in the great majority, as depicted in Fig.~\ref{fig:RS-state}.
It was then conjectured~\cite{Fisher1994} that the quantity $r^{-\psi}\overline{\ln\left|\left\langle S_{i}^{\alpha}S_{i+r}^{\alpha}\right\rangle \right|}$
converges to a distance-independent distribution. Therefore, 
\begin{equation}
C_{\text{typ}}^{\alpha\alpha}\left(r\right)\equiv\exp\overline{\ln\left|\left\langle S_{i}^{\alpha}S_{i+r}^{\alpha}\right\rangle \right|}\sim\exp\left(-\text{const}\times r^{\psi}\right),\label{eq:typ-C}
\end{equation}
with universal and isotropic tunneling exponent $\psi=\frac{1}{2}$.
This result was confirmed in Ref.~\cite{Henelius1998} but its dependence
with the disorder strength (encoded in the constant prefactor) remains
unknown.

\subsection{Some known results for the case of correlated couplings}

In contrast, for the case of locally correlated couplings (the sequence
of couplings being \{$J_{1},J_{1},J_{2},J_{2},\dots J_{\frac{L}{2}},J_{\frac{L}{2}}$\})
and anisotropy parameter $\Delta=0$, the physics is quite different~\cite{Hoyos2011,Getelina2016,Getelina2018}.

For weak disorder $D<D_{c}\approx0.3$, the critical properties are
those of the clean system, i.e., weak disorder is an irrelevant perturbation.
Hence, the mean and typical values of the correlation functions are
approximately equal, and the corresponding exponents are those of
the clean system, i.e., $C^{\alpha\alpha}\approx C_{\text{clean}}^{\alpha\alpha}\sim r^{-\eta_{\alpha}}$,
with $\eta_{x}=\frac{1}{2}$ and $\eta_{z}=2$.

For $D>D_{c}$, a line of finite-disorder fixed points is tuned and
thus the critical exponents vary continuously with the disorder strength~\cite{Hoyos2011}.
However, in contrast with the infinite-randomness case, we only know
that the longitudinal mean correlations decays algebraically with
apparently disorder-independent exponent $\eta_{z}\approx2$~\cite{Getelina2016}.

We recall that the effects of long-range correlated disorder in closely
related systems have been studied in Refs.~\onlinecite{rieger-igloi-prl99,IBV-pre14}.
It was shown that disorder effects are actually enhanced, i.e., the
critical theory is of infinite-randomness type accompanied with offcritical
enhanced Griffiths singularities. We stress that our correlated disorder
has the quite opposite effect~\cite{Hoyos2011}.

\subsection{Methods and further motivations\label{subsec:Methods}}

One of our main goals is to study the non-universal numerical prefactors
of the correlation functions. As there is no analytical theory capable
of dealing with the clean and random systems on the same footing,
we then resort to exact diagonalization of large systems. This is
possible only for the $\Delta=0$ case via the mapping of the Hamiltonian
\eqref{eq:H} into free spinless fermions~\cite{Lieb1961}.

Nonetheless, this is not as simple as it looks. Due to the singularities
of strongly disordered systems (namely, large dynamical exponent),
we had to use quadruple precision ($32$ decimal places) in the numerical
diagonalization process.

Moreover, regarding the choice of $\Delta=0$, even though it represents
a ``non-interacting'' system, notice it captures the universal infinite-randomness
quantum critical properties (as predicted by the SDRG method) of the
entire $-\frac{1}{2}<\Delta<1$ line, i.e., interactions are RG irrelevant
in this range~\cite{Fisher1994}. For the case of correlated couplings,
studying the $\Delta=0$ case is imperative since the finite-disorder
character can only be explored for $\Delta=0$~\cite{Getelina2016}.

Finally, given that the SDRG method is believed to provide exact results
concerning the critical singularities of the model \eqref{eq:H},
it is desirable to investigate large system sizes in order to check
the logarithmic corrections mentioned in Eqs.~\eqref{eq:mean-Cx-Igloi}
and \eqref{eq:mean-C-Sandvik}. The motivation for searching them
is justified in the early works of homogeneous XXZ spin-$1/2$ chains~\cite{Affleck1989,Giamarchi1989,Singh1989,Hallberg1995,Affleck1998,Lukyanov1998,Hikihara1998},
and also in a recent work of the random XXZ model at $\Delta=-\frac{1}{2}$~\cite{Giamarchi2012}.
We anticipate that our results are in agreement with \eqref{eq:mean-C}.

\section{Spin-spin correlations for the uncorrelated coupling constants model\label{sec:uncorr}}

We show in this section our results on the (arithmetic average) mean
and (geometric average) typical spin-spin correlation functions in
the ground state of \eqref{eq:H} for $\Delta=0$ and for uncorrelated
disordered coupling constants. We have used quadruple precision ($32$
decimal places) in order to ensure numerical stability.

\subsection{The mean value of the critical correlation function\label{sec:avg}}

We start our study with the mean correlation function. All data here
presented were averaged over $N=10^{6}$ distinct disorder realizations,
except for those cases of system size $L=1\,600$ in which $N=10^{5}$.

\subsubsection{Longitudinal correlations}

\begin{figure}[t]
\begin{centering}
\includegraphics[clip,width=0.8\columnwidth]{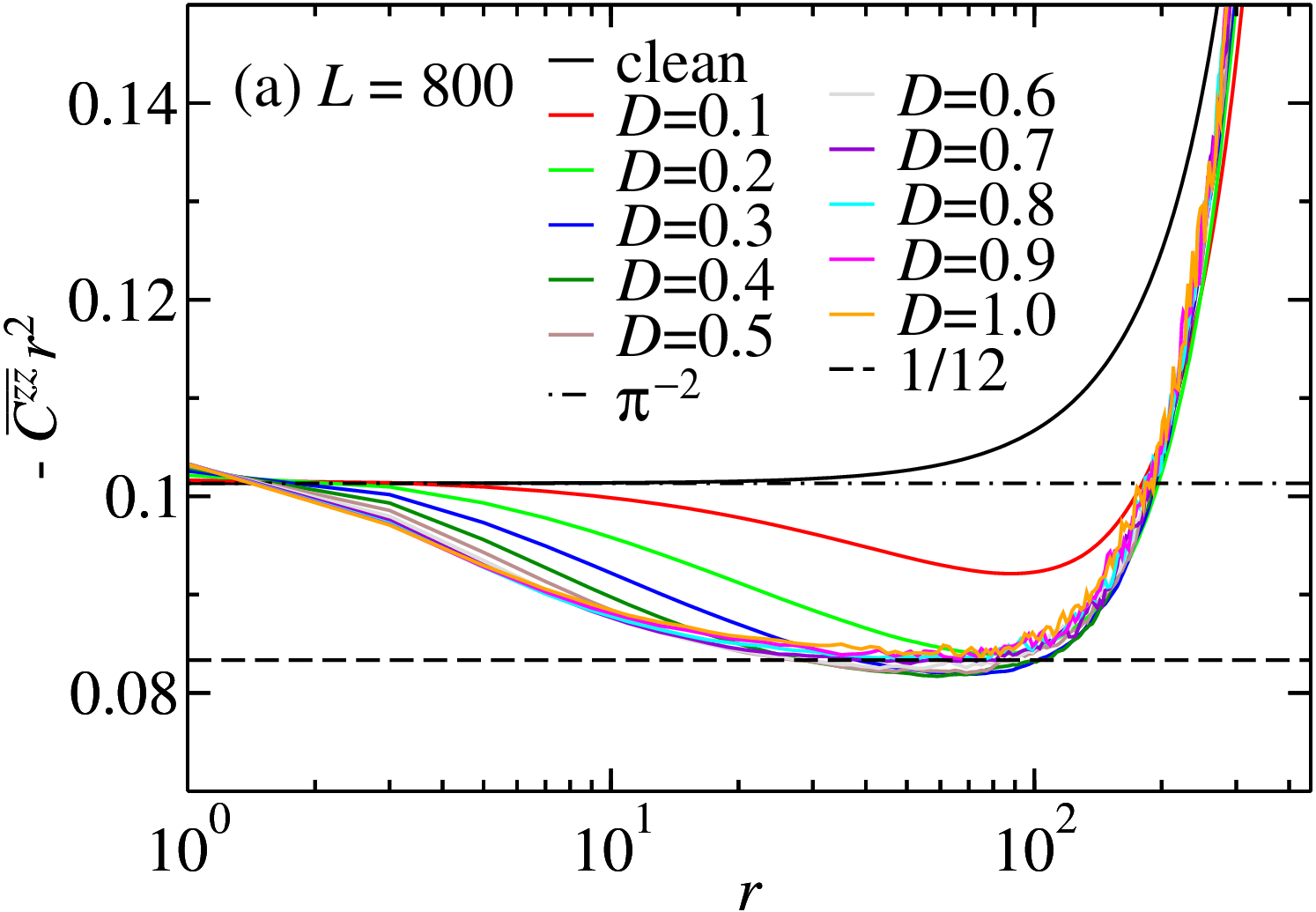}\\
 \includegraphics[clip,width=0.8\columnwidth]{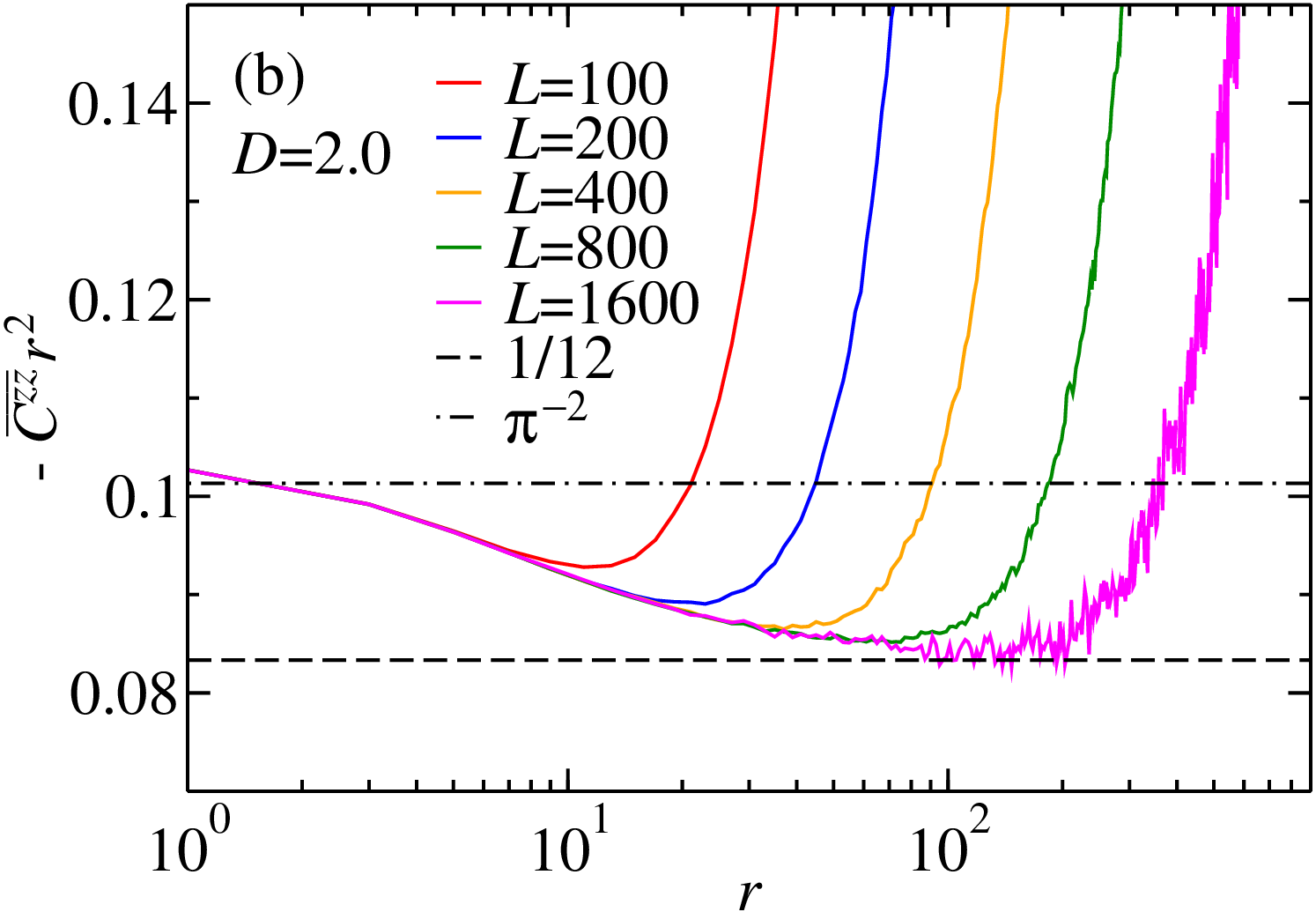}\\
\includegraphics[clip,width=0.8\columnwidth]{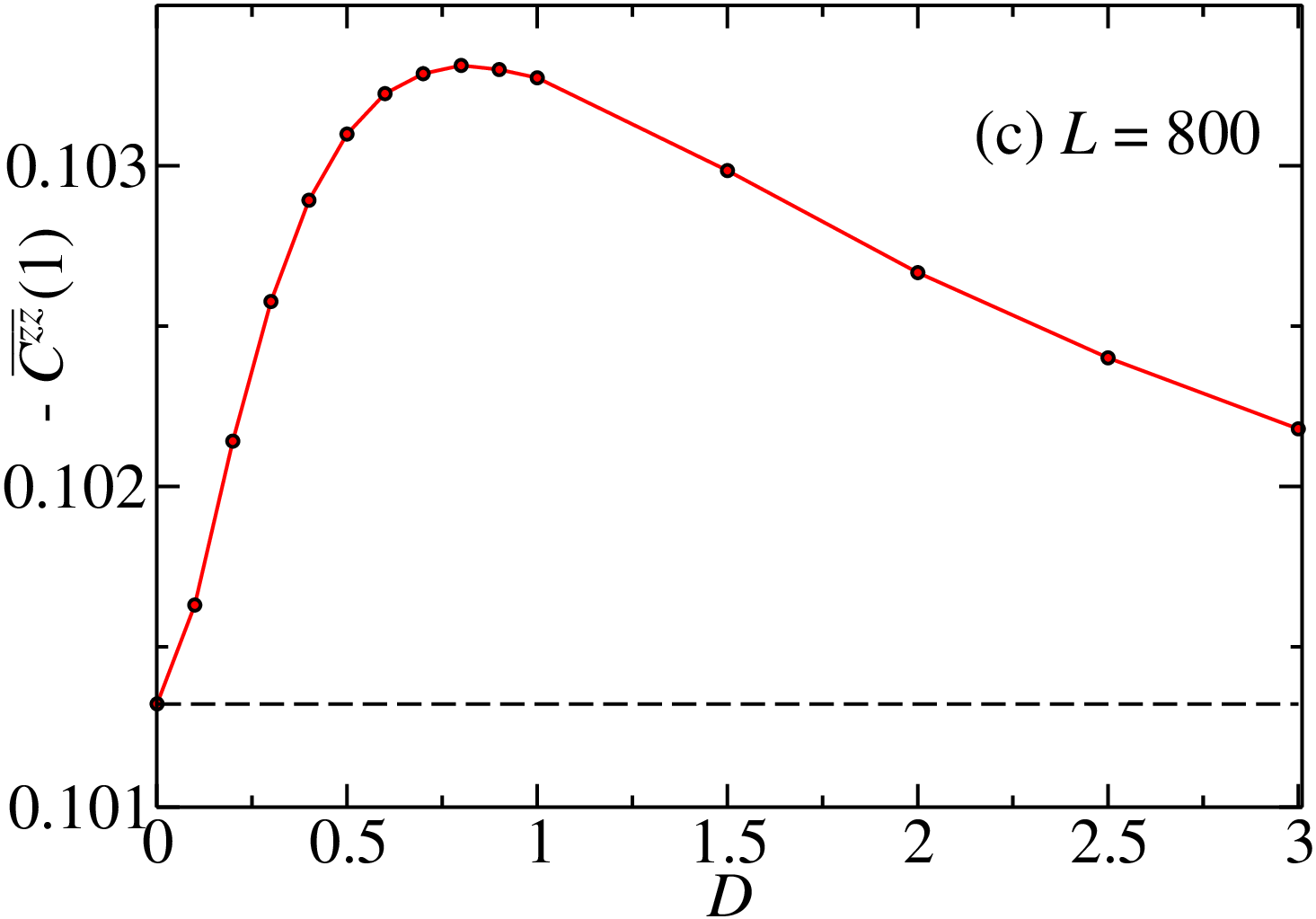}
\par\end{centering}
\caption{\label{fig:zcorr} The mean longitudinal correlation function $\overline{C^{zz}}(r)$
for various chain sizes $L$ and disorder strengths as a function
of the separation $r$ in panels (a) and (b), and $\overline{C^{zz}}(r=1)$
for various $D$ and $L=800$ in panel (c).}
\end{figure}

In Fig.~\ref{fig:zcorr}, we show $\overline{C^{zz}}$ for fixed
system size $L=800$ and various disorder strengths $D$ in panel
\hyperref[fig:zcorr]{(a)}, and fixed $D=2.0$ and various system
sizes $L$ in panel \hyperref[fig:zcorr]{(b)}. The algebraic decay
$\overline{C^{zz}}\sim Ar^{-2}$ is identical in both clean and disordered
case. The difference is in the numerical prefactor: $A=\pi^{-2}$
in the clean case~\cite{Lieb1961}, and is conjectured to be $1/12$
in the disordered case~\cite{Hoyos2007}. As we are interested in
the long-distance behavior $r\gg\xi_{D}$ (but not restricted to $r\ll L$),
with $\xi_{D}$ being a clean-disorder crossover length yet to be
defined, we then assume that the longitudinal correlation function
is 
\begin{equation}
\overline{C^{zz}}\left(r\right)=-\frac{1}{12}\chi_{z}\left(D,r\right)\left(\ell f_{z}\left(\frac{r}{L}\right)\right)^{-\eta},\label{eq:Czz-mean-general}
\end{equation}
 where $\eta=2$, 
\begin{equation}
\ell=\frac{L}{\pi}\sin\left(\frac{\pi r}{L}\right),\label{eq:chord-length}
\end{equation}
 is the chord length,~\footnote{If the spins were arranged in a circle of perimeter $L$, then the
chord length $\ell$ is the Euclidean distance between them.} 
\begin{equation}
f_{\alpha}\left(x\right)=1+\sum_{n=1}^{\infty}a_{2n,\alpha}\sin^{2n}\left(\pi x\right),\label{eq:f-correction}
\end{equation}
 (with $\alpha=x$ or $z$) and $\chi_{z}$ is a crossover function
which assumes the value $12\pi^{-2}$ in the small separation limit
($r\ll\xi_{D}$) and converges to $1$ otherwise. From Fig.~\hyperref[fig:zcorr]{\ref{fig:zcorr}(a)},
it clearly converges to $1$ non-monotonically with respect to $D$
and, from Fig.~\hyperref[fig:zcorr]{\ref{fig:zcorr}(b)}, the convergence
happens only after long separations. This non-monotonic behavior can
be also seen in Fig.~\hyperref[fig:zcorr]{\ref{fig:zcorr}(c)} where
the mean correlation for nearest neighbors, $r=1$, is plotted as
a function of $D$ for $L=800.$ Initially it increases (as expected
according to the random singlet picture) but then diminishes for larger
$D$. Evidently, this non-monotonic behavior is related to the total
spin conservation in the $z$ direction. In other words, $\chi_{z}$
is a non-trivial crossover function and will not be studied here.

In the large separation regime $r\gg\xi_{D}$, the main dependence
of $\overline{C^{zz}}$ on $r$ comes as $\ell f_{z}\left(\frac{r}{L}\right)$.
Simply, it is the most generic function consistent with the periodic
boundary conditions: $C\left(r+L\right)=C(r)$ and $C(L-r)=C(r)$;
with $f_{z}$ being simply a correction to the chord length $\ell$:
the true scaling variable in the clean case $C_{\text{clean}}^{zz}=\left(\pi\ell\right)^{-2}$.~\footnote{Corrections to the chord length were reported in the entanglement
entropy as well~\cite{Fagotti2011}.}

Throughout this work, we assume that the coefficients $a_{2n,\alpha}$
are disorder-independent. There is no reason why this should be the
case. Our assumption, however, is compatible with our numerical data.
Nonetheless, due to statistical fluctuations and the lack of knowledge
on $\chi_{\alpha}$, we cannot exclude that $a_{2n,\alpha}$ are indeed
disorder dependent.

\begin{figure}[t]
\begin{centering}
\includegraphics[clip,width=0.8\columnwidth]{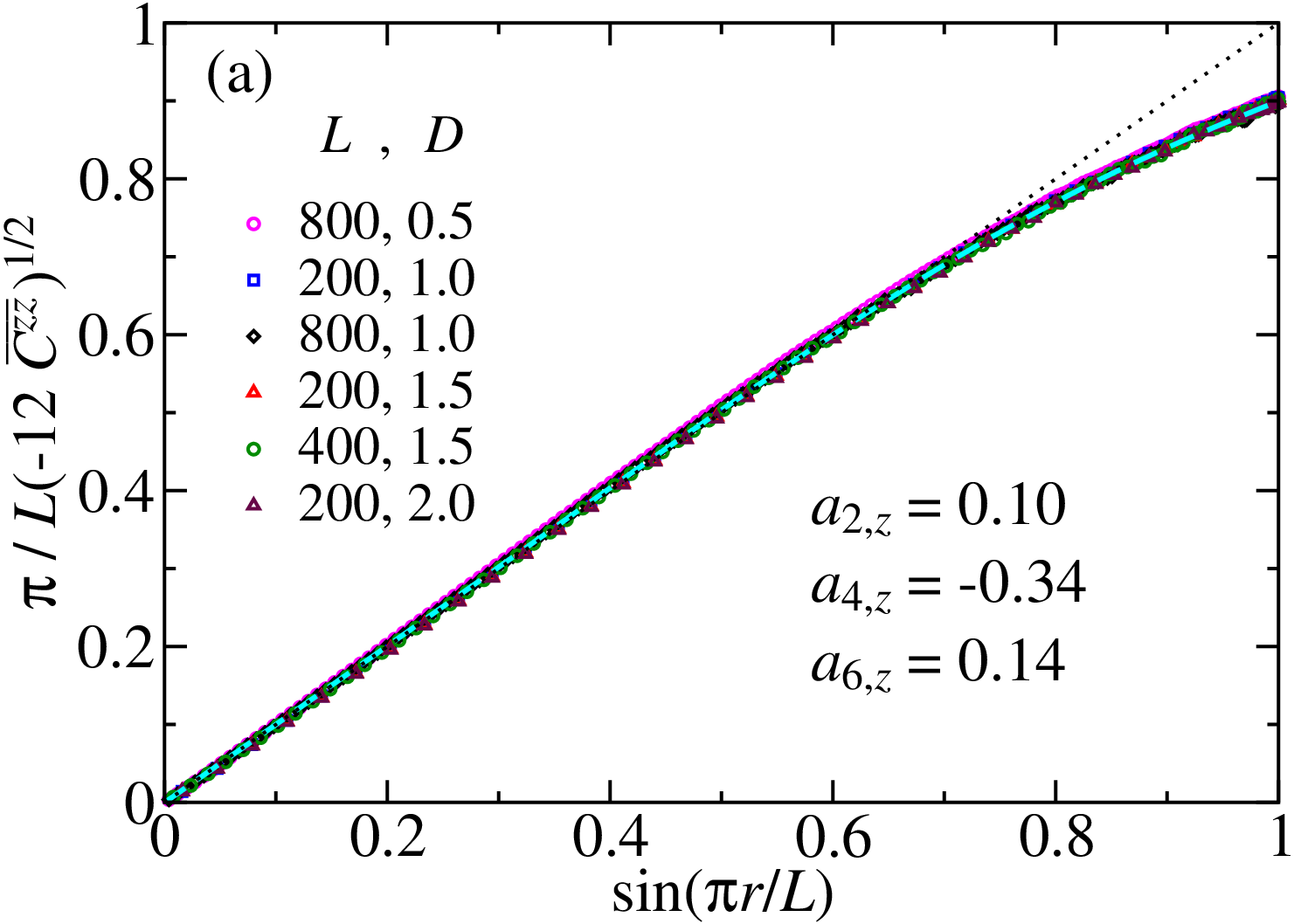}\\
\includegraphics[clip,width=0.8\columnwidth]{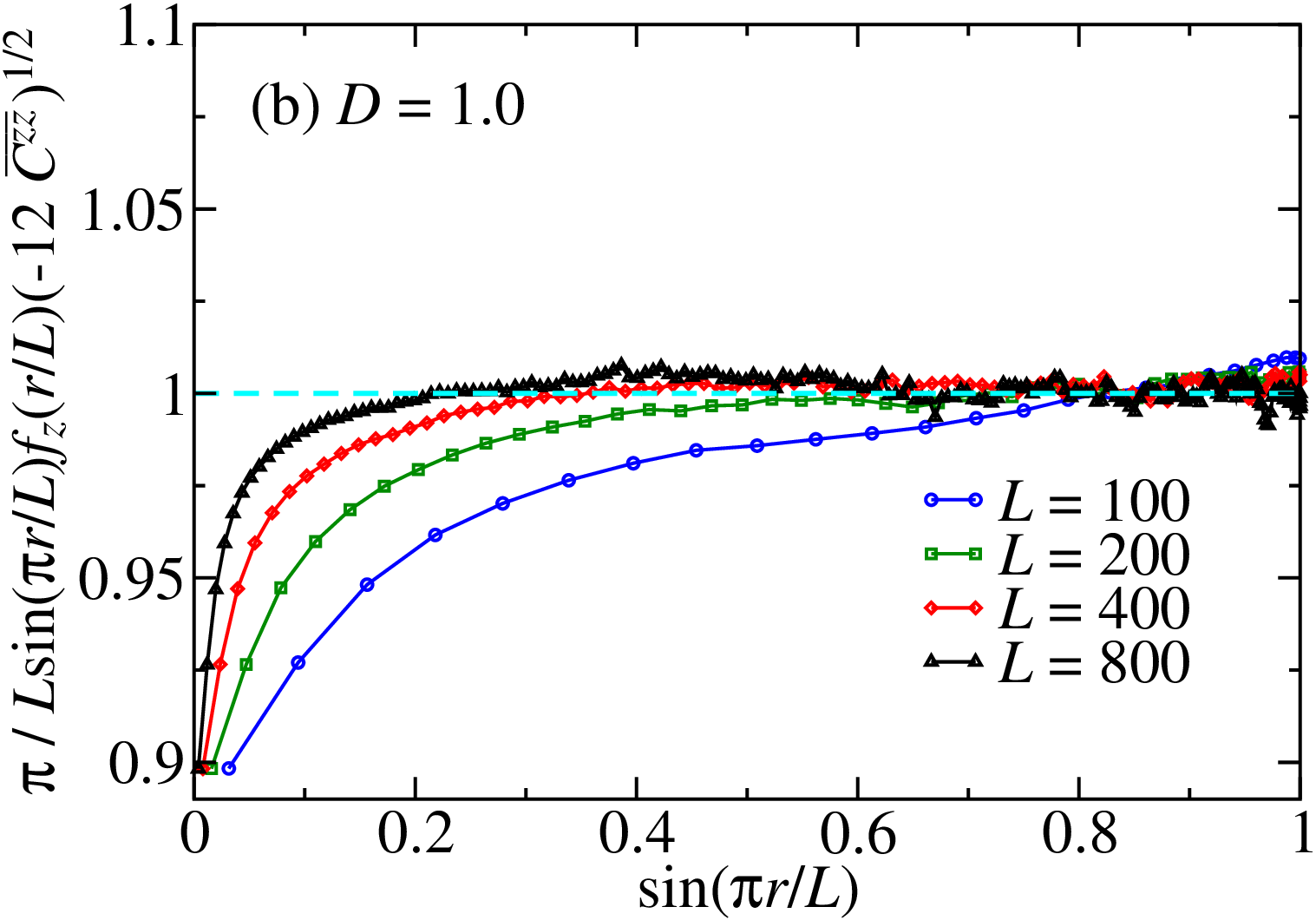}\\
\includegraphics[clip,width=0.8\columnwidth]{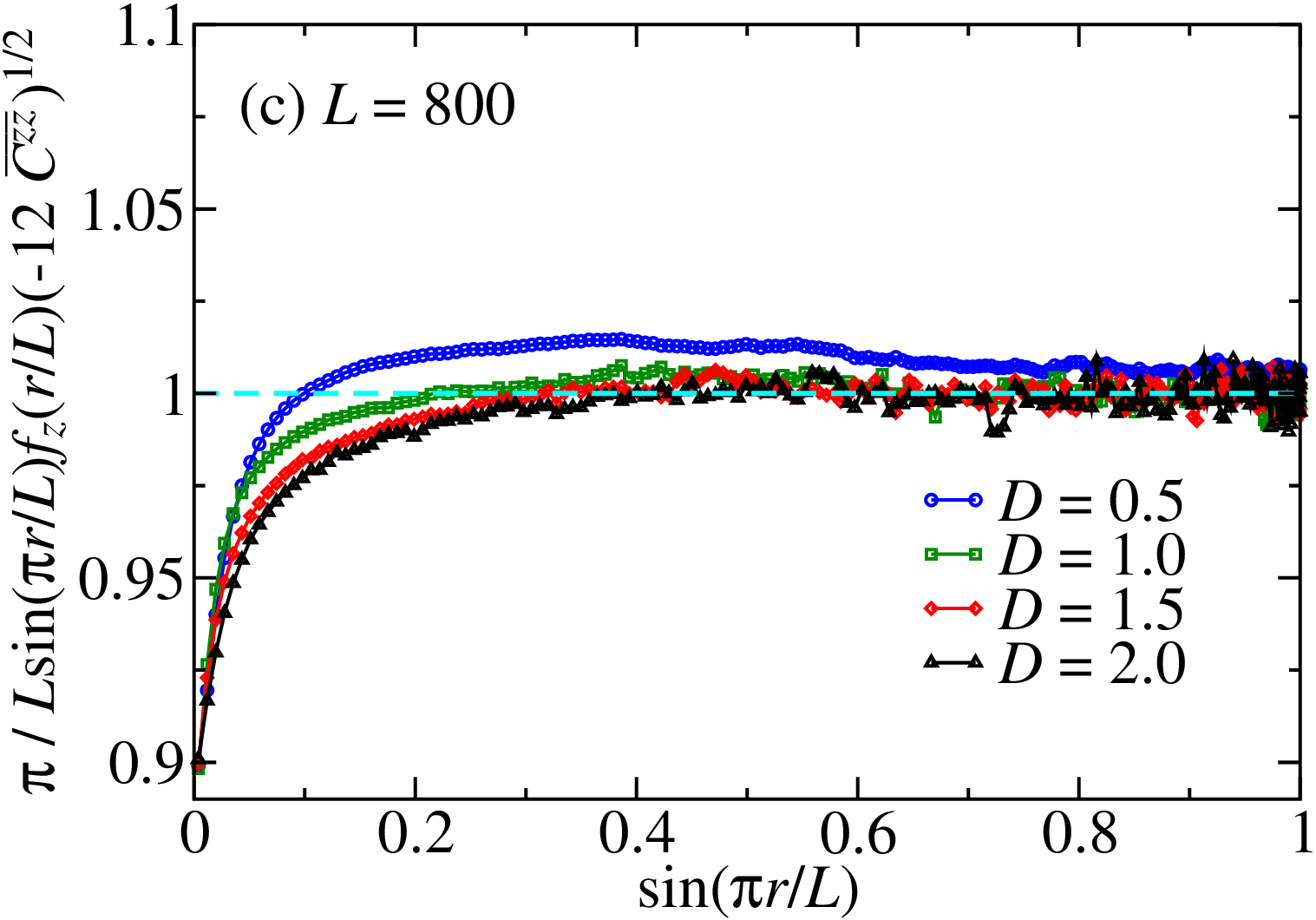}
\par\end{centering}
\caption{\label{fig:czz-fit} The mean longitudinal correlation $\overline{C^{zz}}$
as a function of $\sin\left(\pi r/L\right)$ re-scaled in many different
ways in order to obtain the correction in Eq.~\eqref{eq:f-correction}
(see text). The dashed line in panel (a) corresponds to our best fit:
$a_{2,z}=0.135$, $a_{4,z}=-0.414$ and $a_{6,z}=0.179$. The dotted
line is simply the identity function.}
\end{figure}

In order to obtain the correction to the chord length, we appropriately
replot our data in Fig.~\ref{fig:czz-fit}. In panel \hyperref[fig:czz-fit]{(a)},
we consider only the largest and strongest disordered chains in order
to minimize the effects of the crossover function $\chi_{z}$, i.e.,
we have chosen only systems in which $\chi_{z}$ seems to be very
close to $1$ for a large range of separations $r$. All data collapse
satisfactorily. Tiny deviations are present which, in principle, are
accounted by $\chi_{z}$. From the collapsed data, we then extract
the values of the coefficients $a_{2n,z}$. Best fits using further
corrections (up to $a_{8,z}$) do not improve the reduced weighted
error sum $\bar{\chi}^{2}$. Finally, changing the fitting values
of $a_{2n,z}$ by $10\%$ does not change appreciably the value of
$\bar{\chi}^{2}$, we then estimate that $10\%$ is the accuracy of
our estimates of $a_{2n,z}$.

In Figs.~\hyperref[fig:czz-fit]{\ref{fig:czz-fit}(b)} and \hyperref[fig:czz-fit]{(c)},
we plot the square root of the ratio between $-12\overline{C^{zz}}$
and $\left(\ell f_{z}\left(\frac{r}{L}\right)\right)^{-\eta}$ which
should approach $1$ provided that $f_{z}$ is disorder independent.
In panel \hyperref[fig:czz-fit]{(b)}, disorder strength is fixed
while $L$ is increased. Larger the system size $L$, better the data
is described by the scaling variable $\ell f_{z}\left(\frac{r}{L}\right)$.
Deviations for smaller $L$ are due to the crossover function $\chi_{z}$.
In panel \hyperref[fig:czz-fit]{(c)}, the system size is fixed while
$D$ is changed. Notice the little dependence on $D$ (for large $L$
and the disorder strengths considered). Notice furthermore the non-monotonic
behavior of $\chi_{z}$ with $D$. The convergence to the unity is
faster for intermediate disorder $D\approx1.0$.

\subsubsection{Transverse correlations}

\begin{figure}[t]
\begin{centering}
\includegraphics[clip,width=0.8\columnwidth]{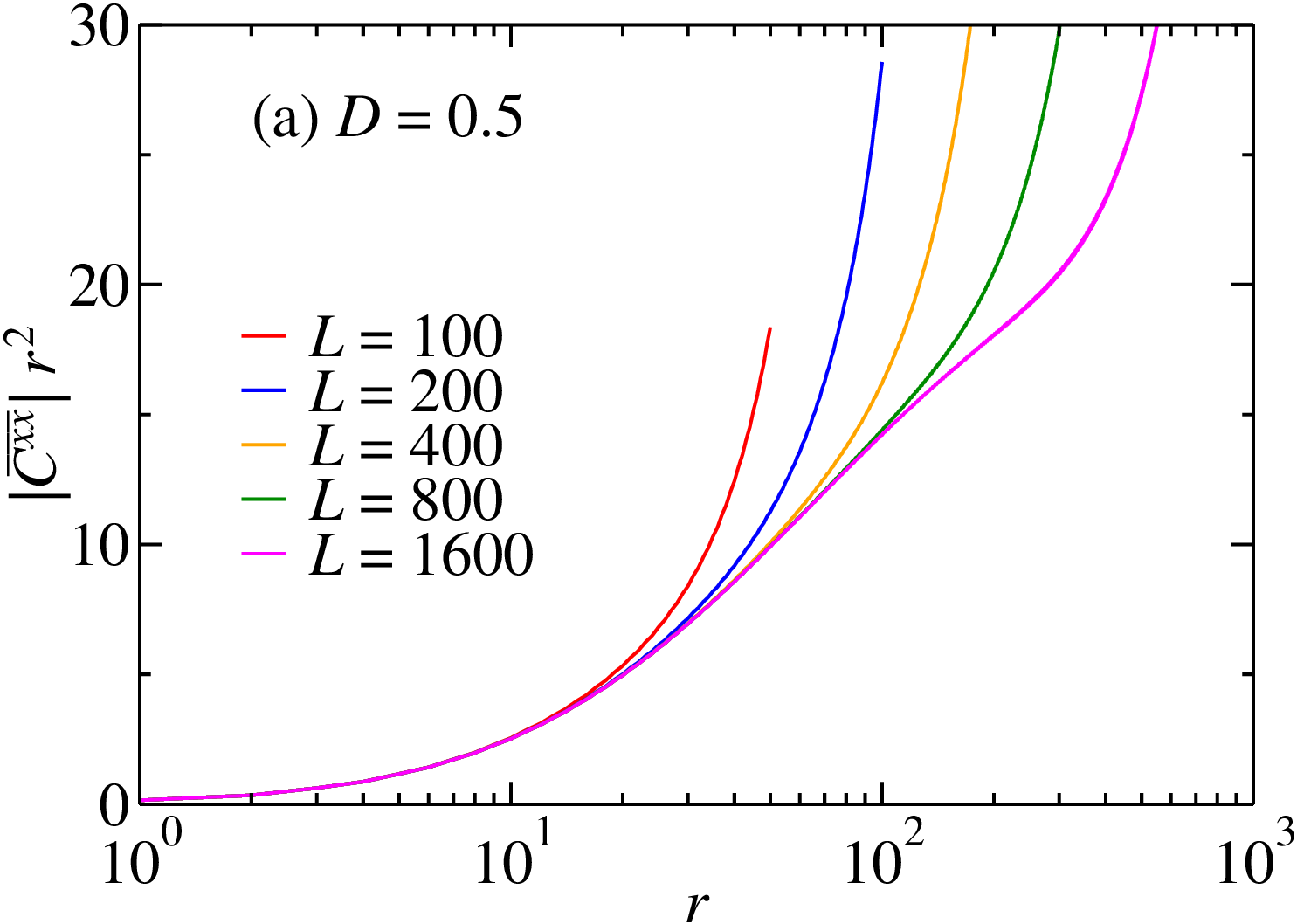}\\
\includegraphics[clip,width=0.8\columnwidth]{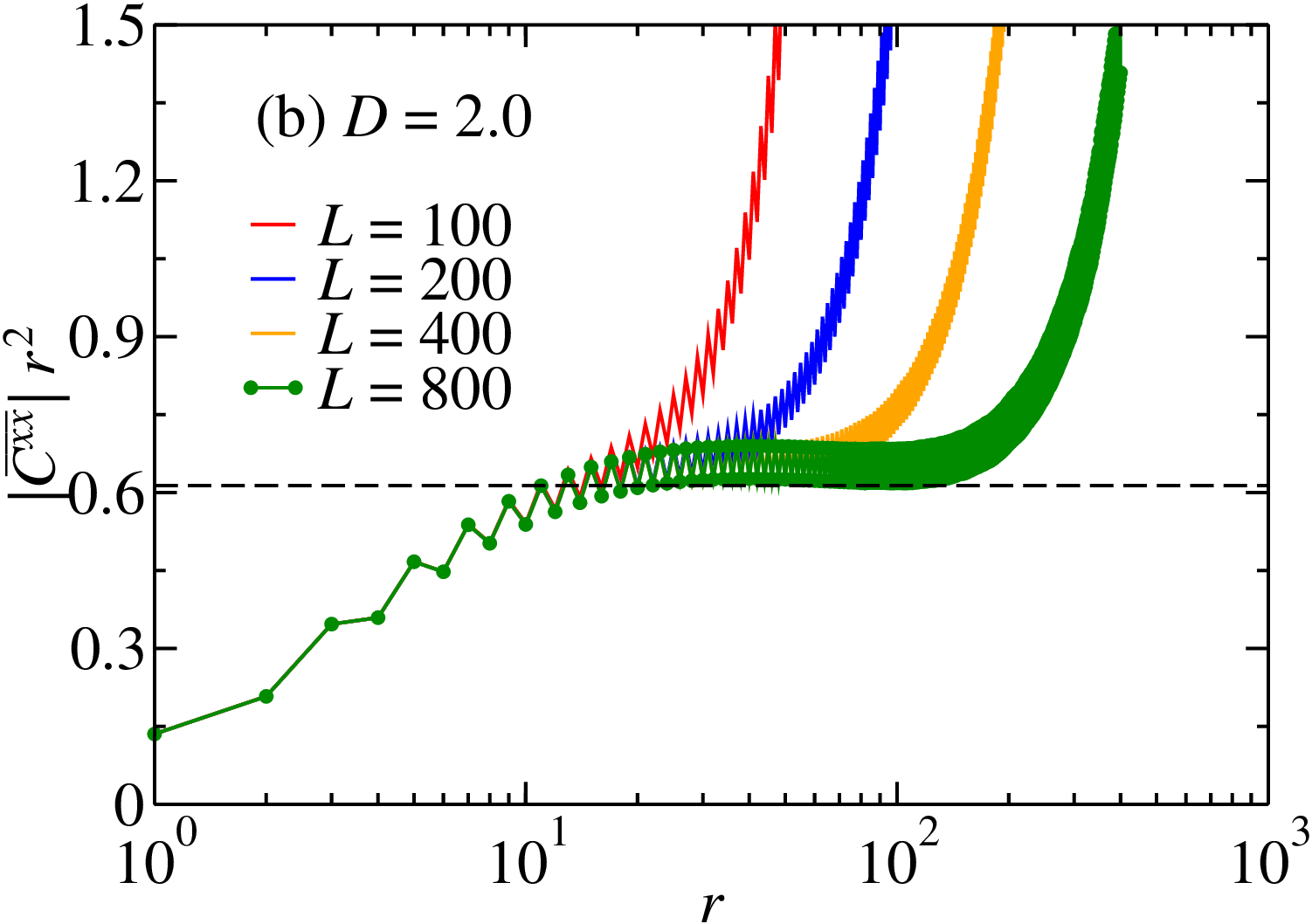} 
\par\end{centering}
\caption{\label{fig:xcorr}The mean transverse correlation function $\left|\overline{C^{xx}}(r)\right|$
for various chain sizes $L$ and disorder strengths (a) $D=0.5$ and
(b) $D=2.0$.}
\end{figure}

The study of the mean transverse correlation function $\overline{C^{xx}}(r)$
is much more involving since (i) it is more numerically demanding,~\footnote{It requires the computation of a large determinant~\cite{Lieb1961}
which makes it more prone to numerical round-off errors.} (ii) there is no knowledge about its numerical prefactor, and, as
shown in Ref.~\onlinecite{Laflorencie2004}, (iii) the clean-disorder
crossover length can be so large that even hinders the clear identification
of the correct algebraic decay exponent $\eta=2$ (see Fig.~\ref{fig:xcorr}).
Moreover and interestingly, as clearly seen in Fig.~\hyperref[fig:xcorr]{\ref{fig:xcorr}(b)},
this numerical prefactor is different from odd and even separations
$r$.~\footnote{In the clean case~\cite{mccoy-68}, the prefactor $\approx0.147\,09$
and is the same for both even and odd separations.}

As for the longitudinal correlations \eqref{eq:Czz-mean-general},
the natural choice for the mean transverse correlation function is
\begin{equation}
\overline{C^{xx}}(r)=\left(-1\right)^{r}c_{D,r}\chi_{x}\left(D,r\right)\left(\ell f_{x}\left(\frac{r}{L}\right)\right)^{-\eta},\label{eq:Cxx-general}
\end{equation}
 where $\eta=2$, and $f_{x}$ is analogous to $f_{z}$ in Eq.~\ref{eq:f-correction}.
Likewise, the crossover function $\chi_{x}$ is expected to be analogous
to $\chi_{z},$ and thus, is a non-trivial function which should be
proportional to $\sim r^{3/2}$ in the $r\ll\xi_{D}$ regime, and
converges to $1$ otherwise. Here, $c_{D,r}$ represents the numerical
prefactor which, in the large separation limit, equals to 
\begin{equation}
c_{D,r}=\frac{1}{24}\left(c_{\text{o},D}+c_{\text{e},D}-\left(-1\right)^{r}\left(c_{\text{o},D}-c_{\text{e},D}\right)\right),\label{eq:prefactor-Cxx}
\end{equation}
 with $c_{\text{o(e)},D}$ being the absolute value of the prefactor
corresponding to odd (even) separations (multiplied by $12$, for
comparison with $\overline{C^{zz}}$).

In order to obtain the chord-length correction $f_{x}$, it is helpful
to have some knowledge of the prefactor $c_{D,r}$. Naively, one could
obtain its dependence with $D$ by simply connecting the clean and
disordered behaviors, i.e., given that $C_{\text{clean}}^{xx}=A/\sqrt{r}$
and that $\overline{C^{xx}}=c_{D,r}/r^{2}$, then $C_{\text{clean}}^{xx}=\overline{C^{xx}}$
at, say, a sharp crossover length $r=\xi_{D}$. Hence, $c_{D,r}\sim\xi_{D}^{3/2}$
and thus, we need knowledge on the crossover length.

Using field-theory methods (accurate at the weak-disorder limit $D\ll1$),
it was shown that $\xi_{D}\sim1/\text{var}\left(J\right)=D^{-2}\left(1+D\right)^{2}\left(1+2D\right)$.
However, while this relation is accurate for small $D$, it was numerically
found that $\xi_{D}\sim D^{-\left(2.0\pm0.2\right)}$ is much more
satisfactory for any $D$~\cite{Laflorencie2004}. Later, it was
shown that a single-parameter theory holds at the band center of particle-hole
symmetric tight-binding chains~\cite{Hoyos2014} {[}which maps to
the Hamiltonian \eqref{eq:H}{]}. The wavefunction is stretched-exponentially
localized with the inverse of the Lyapunov exponent (or localization
length) being~\footnote{Comparing the definition of the Lyapunov exponent \eqref{eq:Lyapunov-exponent}
with the values of the crossover length numerically provided in Ref.~\onlinecite{Laflorencie2004},
we simply find that $\xi_{D}\approx51\gamma_{D}^{-1}\approx20D^{-2}$.}
\begin{equation}
\gamma_{D}^{-1}=\frac{\pi}{8\text{Var}\left(\ln J\right)}=\frac{\pi}{8D^{2}}.\label{eq:Lyapunov-exponent}
\end{equation}

\begin{figure}[b]
\begin{centering}
\includegraphics[clip,width=0.8\columnwidth]{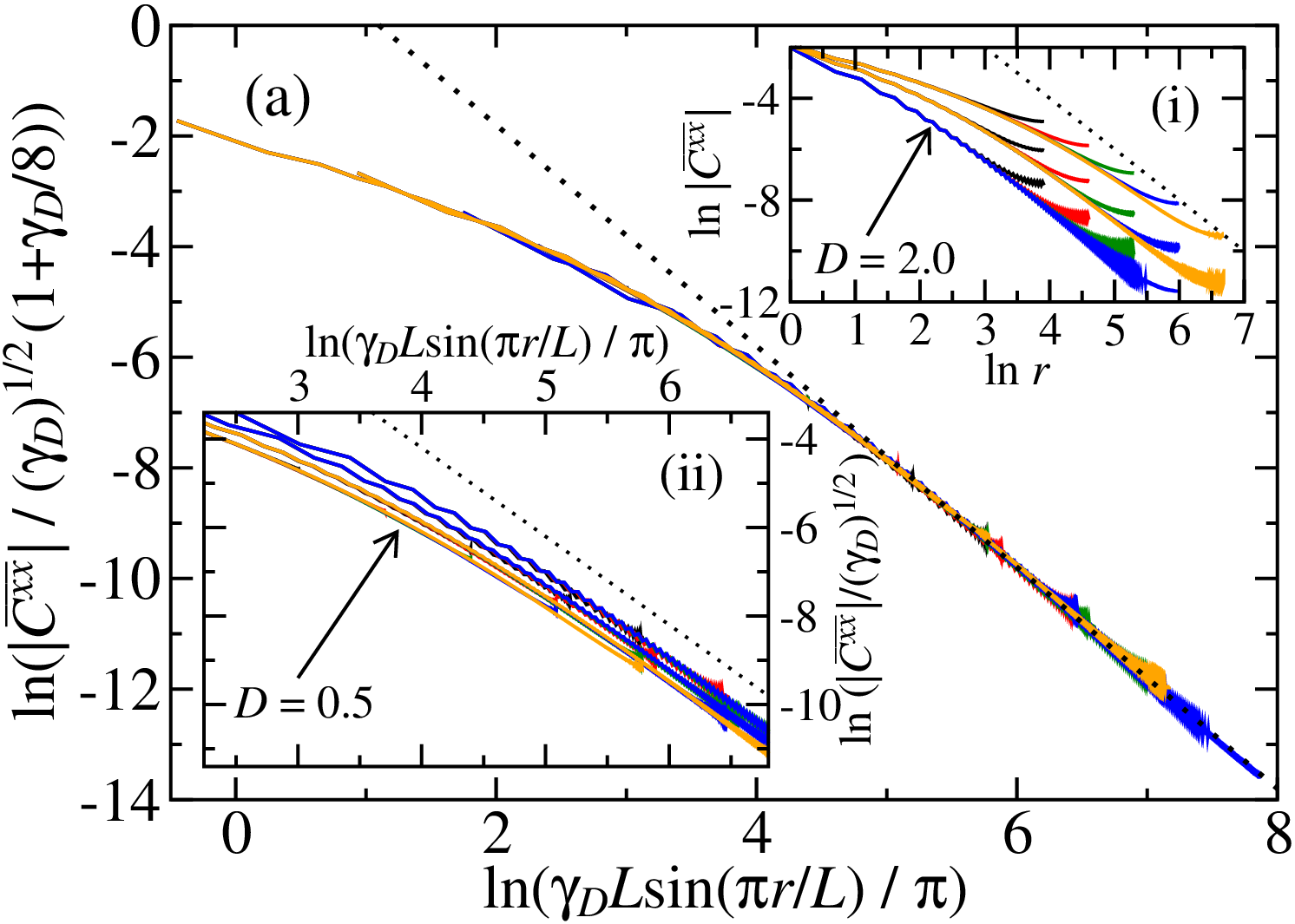}\\
\includegraphics[clip,width=0.8\columnwidth]{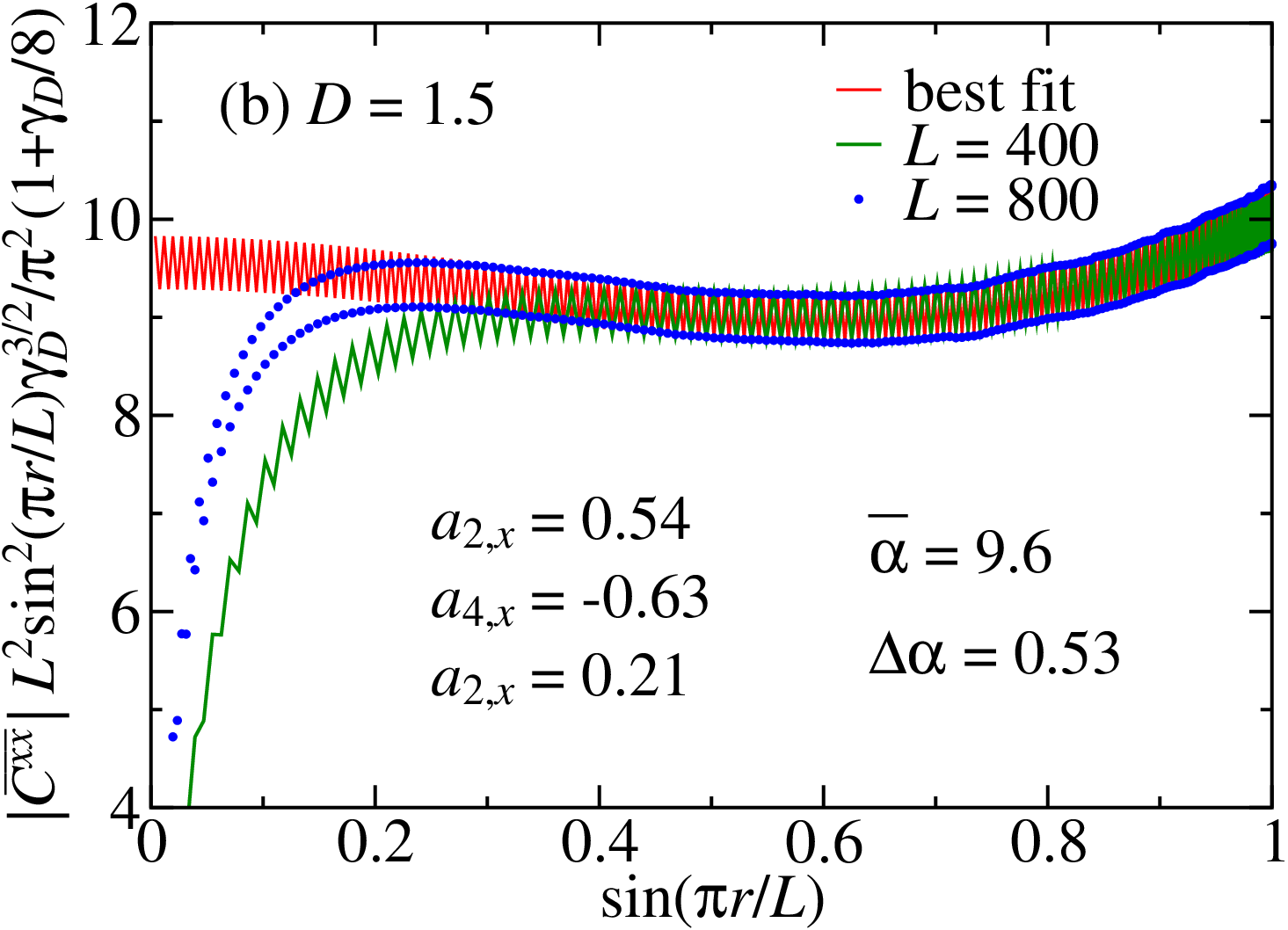} 
\par\end{centering}
\caption{\label{fig:Cxx}(a) The mean transverse correlation function $\overline{C^{xx}}$
as a function of the spin separation $r$ for various disorder strengths
$D=0.5$, $D=1.0$, $D=1.5$ and $D=2.0$ and system sizes $L=100$
(black), $200$ (red), $400$ (green), $800$ (blue) and $1600$ (orange,
and only for $D=0.5$ and $1.0$). For clarity, the data for $D=1.5$
is not shown in inset (i). The dotted lines are $y\sim x^{-2}$ for
comparison. (b) The data is rescaled in order to highlight the correction
to the chord length scaling (see text). In addition, the numerical
prefactors in\eqref{eq:prefactor-Cxx2} can be extracted via a fitting
and are $\bar{\alpha}=9.6(2)$, $\Delta\alpha_{D=0.5}=0.018(3)$,
$\Delta\alpha_{D=1.0}=0.18(1)$, $\Delta\alpha_{D=1.5}=0.53(1)$,
and $\Delta\alpha_{D=2.0}=1.06(2)$.}
\end{figure}

With those arguments in mind, we now try to rescale the $\overline{C^{xx}}$
{[}shown in the inset (i) of Fig.~\hyperref[fig:Cxx]{\ref{fig:Cxx}(a)}{]}
appropriately. Given that (i) $c_{D,r}\sim\gamma_{D}^{-3/2}$ (naive
crossover), that (ii) the natural length scale is $\gamma_{D}^{-1}$
and that (iii) the chord length $\ell$ is weakly corrected, we then
rescale the chord length in units of $\gamma_{D}^{-1}$ and, therefore,
$\overline{C^{xx}}$ must scale as $\sim\sqrt{\gamma_{D}}$. Somewhat
surprisingly, with this naive rescaling {[}see inset (ii) of Fig.~\hyperref[fig:Cxx]{\ref{fig:Cxx}(a)}{]}
we almost achieve a perfect data collapse. In order to improve the
data collapse, we fit the data in the inset (ii) to a power-law function
$A\left(\gamma_{D}\ell\right)^{-2}$, and find that $A\propto1+0.125\gamma_{D}$
in the long distance regime $\gamma_{D}\ell\gg1$. We then correct
our naive scaling to 
\begin{equation}
\overline{C^{xx}}\sim\left(1+\frac{1}{8}\gamma_{D}\right)\sqrt{\gamma_{D}}
\end{equation}
 and plot the resulting data in the main panel of Fig.~\hyperref[fig:Cxx]{\ref{fig:Cxx}(a)}.
The collapse is remarkable even for small separations, suggesting
a crossover function $\chi_{x}\left(D,r\right)\approx\chi_{x}\left(\gamma_{D}r\right)$
for $\gamma_{D}r\apprge1$. In addition, we find useful to recast
the prefactor \eqref{eq:prefactor-Cxx} as 
\begin{equation}
c_{D,r}=\left(1+\frac{1}{8}\gamma_{D}\right)\gamma_{D}^{-3/2}\left(\bar{\alpha}-\frac{1}{2}\left(-1\right)^{r}\Delta\alpha_{D}\right),\label{eq:prefactor-Cxx2}
\end{equation}
 where $\bar{\alpha}$ is disorder independent.

In Fig.~\hyperref[fig:Cxx]{\ref{fig:Cxx}(b)}, we plot $Y=\overline{C^{xx}}\left(\gamma_{D}\ell\right)^{2}/\left(1+\frac{1}{8}\gamma_{D}\right)\sqrt{\gamma_{D}}$
as a function of $X=\sin\left(\pi r/L\right)$ in order to obtain
the values $a_{2n,x}$, $\bar{\alpha}$ and $\Delta\alpha_{D}$. This
is achieved via, according to \eqref{eq:Cxx-general}, fitting $Y=\chi_{x}\left(\bar{\alpha}-\frac{1}{2}\left(-1\right)^{r}\Delta\alpha_{D}\right)/f_{x}(X)$
to our data. For clarity, we have shown only the data for $L=400$
and $800$ and $D=1.5$.~\footnote{We report that the corresponding curves for $D=0.5$, $1.0$ and $2.0$
are quite consistent with the collapse in \hyperref[fig:Cxx]{\ref{fig:Cxx}(a)}.
The only difference being on the value of $\Delta\alpha_{D}$. For
$D=0.5,$ the crossover function $\chi_{x}$ is evidently larger yielding
a smaller fitting region $X\apprge0.8$.} Clearly, the crossover function $\chi_{x}$ is converged to $1$
for $X\apprge0.4$ (our fitting region) and $L=800$. The best fit
is shown as a solid red line. Notice that this plot is similar to
those in panels (b) and (c) of Fig.~\ref{fig:czz-fit}. 

Finally, it is interesting to observe the prefactor difference $\Delta c_{D}=c_{\text{o},D}-c_{\text{e},D}$
and mean value $\bar{c}_{D}=\frac{1}{2}\left(c_{\text{o},D}+c_{\text{e},D}\right)$.
Using the relations \eqref{eq:prefactor-Cxx} and \eqref{eq:prefactor-Cxx2},
together with the values of $\Delta\alpha_{D}$ and $\bar{\alpha}$
listed in the caption of Fig.~\ref{fig:Cxx}, we find that $\left(\Delta c_{D},\bar{c}_{D}\right)=\left(0.46,244\right)$,
$\left(0.70,37\right)$, $\left(0.80,14\right)$, and $\left(0.89,8\right)$
for $D=0.5$, $1.0$, $1.5$, and $2.0$, respectively. Notice it
is not much different from $1$ for all values of $D$. As conjectured
in Ref.~\cite{Hoyos2007}, this difference should be equal $1$ for
correlations along a symmetry axis. The total magnetization in the
$x$ direction is not conserved. However, perhaps due to the the emergent
symmetry SO(2)$\rightarrow$SU(2) character of the random singlet
state, violations of this difference are small when compared to the
values of the coefficients themselves.

\subsubsection{Mimicking logarithmic corrections}

\begin{figure}[t]
\begin{centering}
\includegraphics[clip,width=0.8\columnwidth]{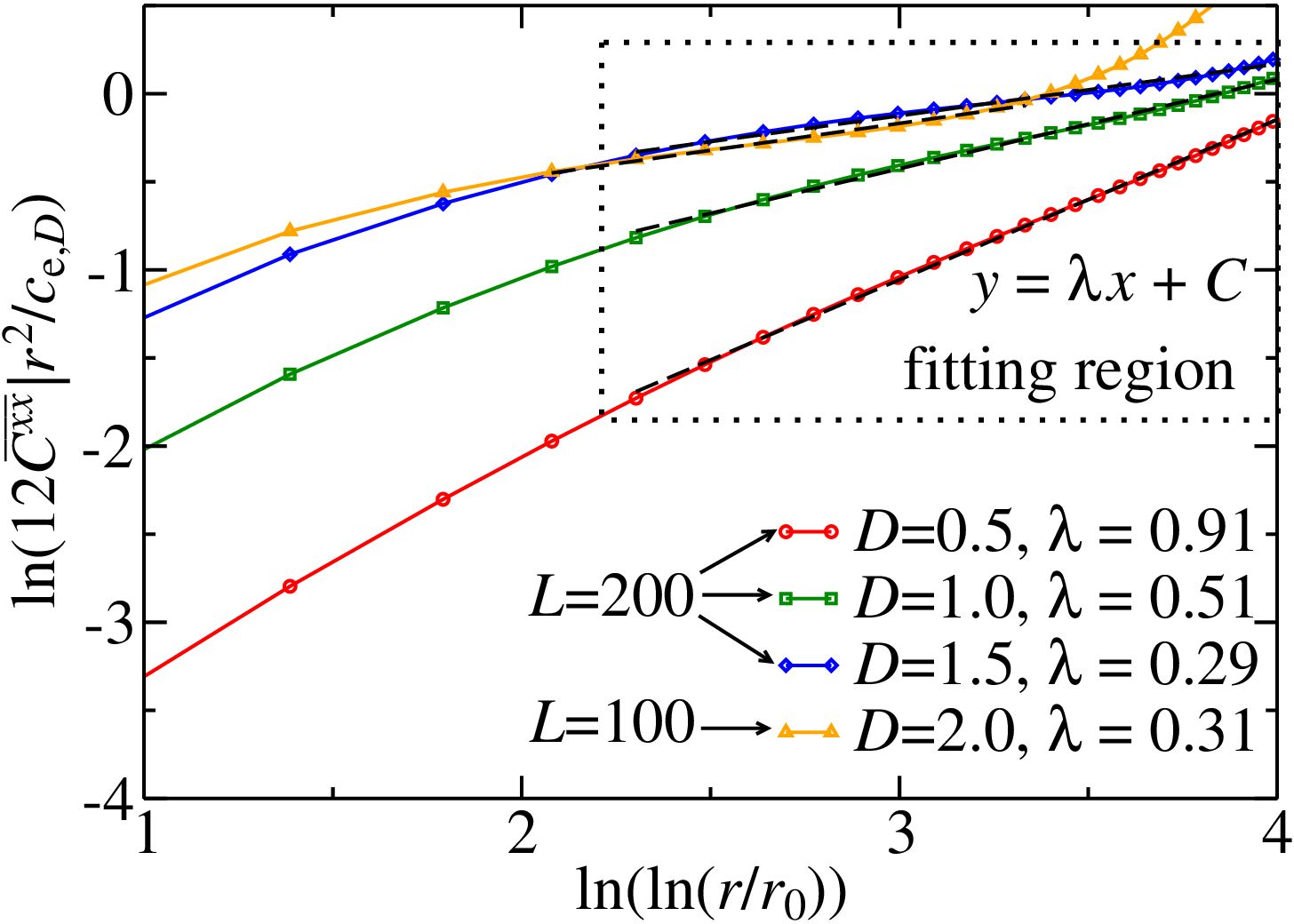}
\par\end{centering}
\caption{\label{fig:sqrt}Mean transverse correlation function $\overline{C^{xx}}(r)$
plotted according to Eq.~\eqref{eq:fit-log} for even separations
$r$, considering various disorder strengths $D$ and chain sizes
$L$. The upturn for $L=100$ is due to the periodic boundary condition.
For comparison, the data is fitted to the function $y=\lambda x+\text{const}$
(dashed lines) within the region compatible with that of Ref.~\onlinecite{Shu2016}.}
\end{figure}

Having characterized the long-distance ($\gamma_{D}r\gg1$) behavior
of the transverse mean correlation function Eq.~\eqref{eq:Cxx-general},
we now call attention to their strong finite-size effects when characterizing
the random-singlet state in numerical studies via the use of small
systems. Interestingly, as can be seen in Fig.~\ref{fig:sqrt}, the
data is compatible with a logarithmic correction to the SDRG prediction
of the leading term, i.e., based on small system sizes, one could
conclude that 
\begin{equation}
\overline{C^{xx}}(r)r^{2}\sim\ln^{\lambda}\left(\frac{r}{r_{0}}\right).\label{eq:fit-log}
\end{equation}
 Corrections to the SDRG prediction were reported in the literature
over the years, ranging from non-universal critical exponents~\cite{hamacher-etal-prl02}
to logarithmic corrections~\cite{Igloi2000,Shu2016}. Here, we have
plotted and fitted our data in the same range compatible with those
of Ref.~\onlinecite{Shu2016} for the Heisenberg random spin chain.
The values of the corresponding effective exponent $\lambda$ are
within the range found in that work. 

We emphasize that our data show no evidence of logarithmic corrections
in the long distance and long chain regime. We also stress that we
do not have shown the absence of the logarithmic corrections reported
in Ref.~\onlinecite{Shu2016} (which study the isotropic $\Delta=1$
case), we have only showed that the combination of crossover and finite-size
effects {[}as in Eq.~\eqref{eq:Cxx-general}{]} can be interpreted
as logarithmic corrections in the case of the XX random spin-1/2 chain.
The main culprit being the crossover function $\chi_{x}$.

\subsubsection{Random-singlet correlations}

\begin{figure}[t]
\begin{centering}
\includegraphics[clip,width=0.8\columnwidth]{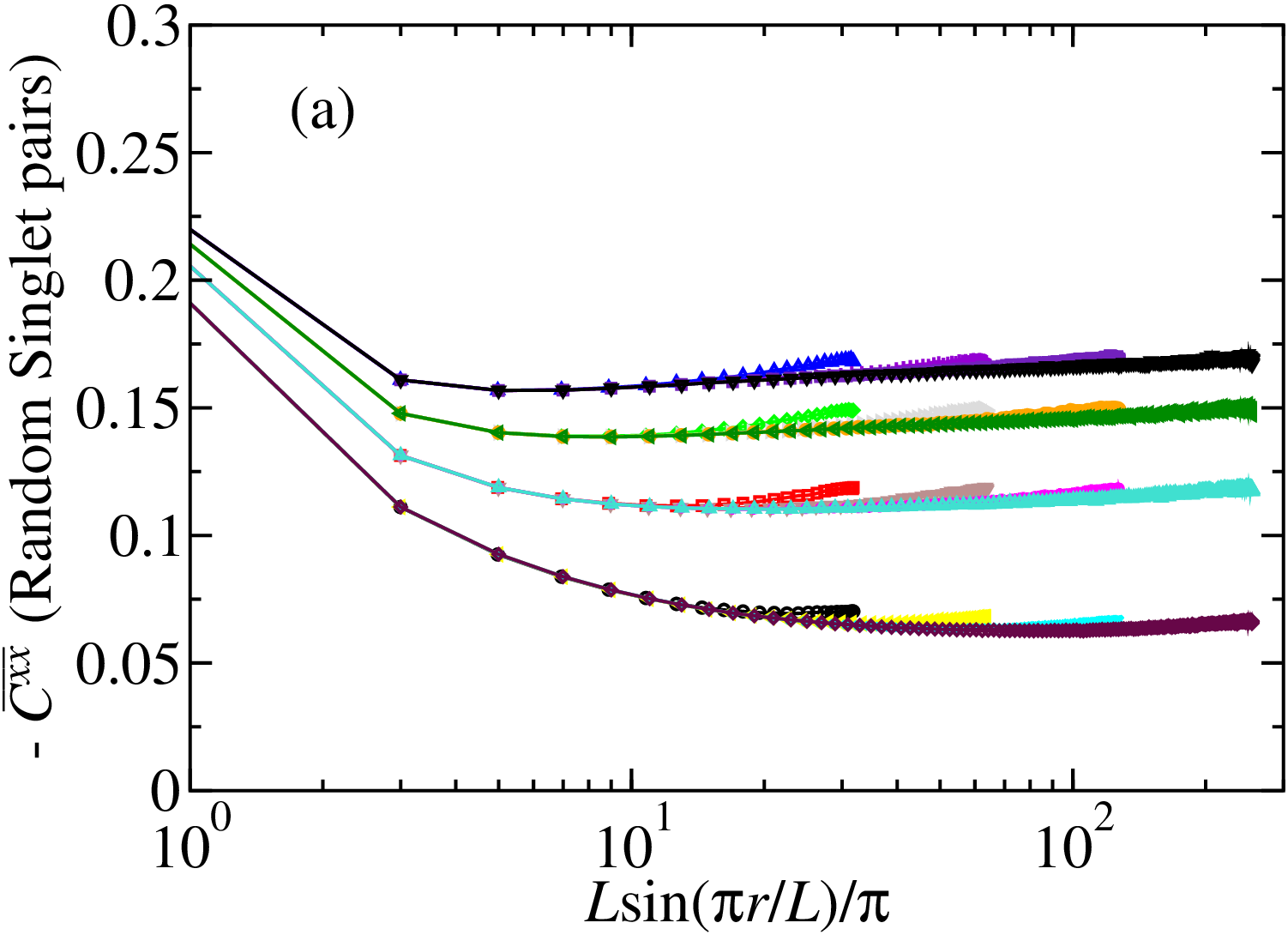} \\
\includegraphics[clip,width=0.8\columnwidth]{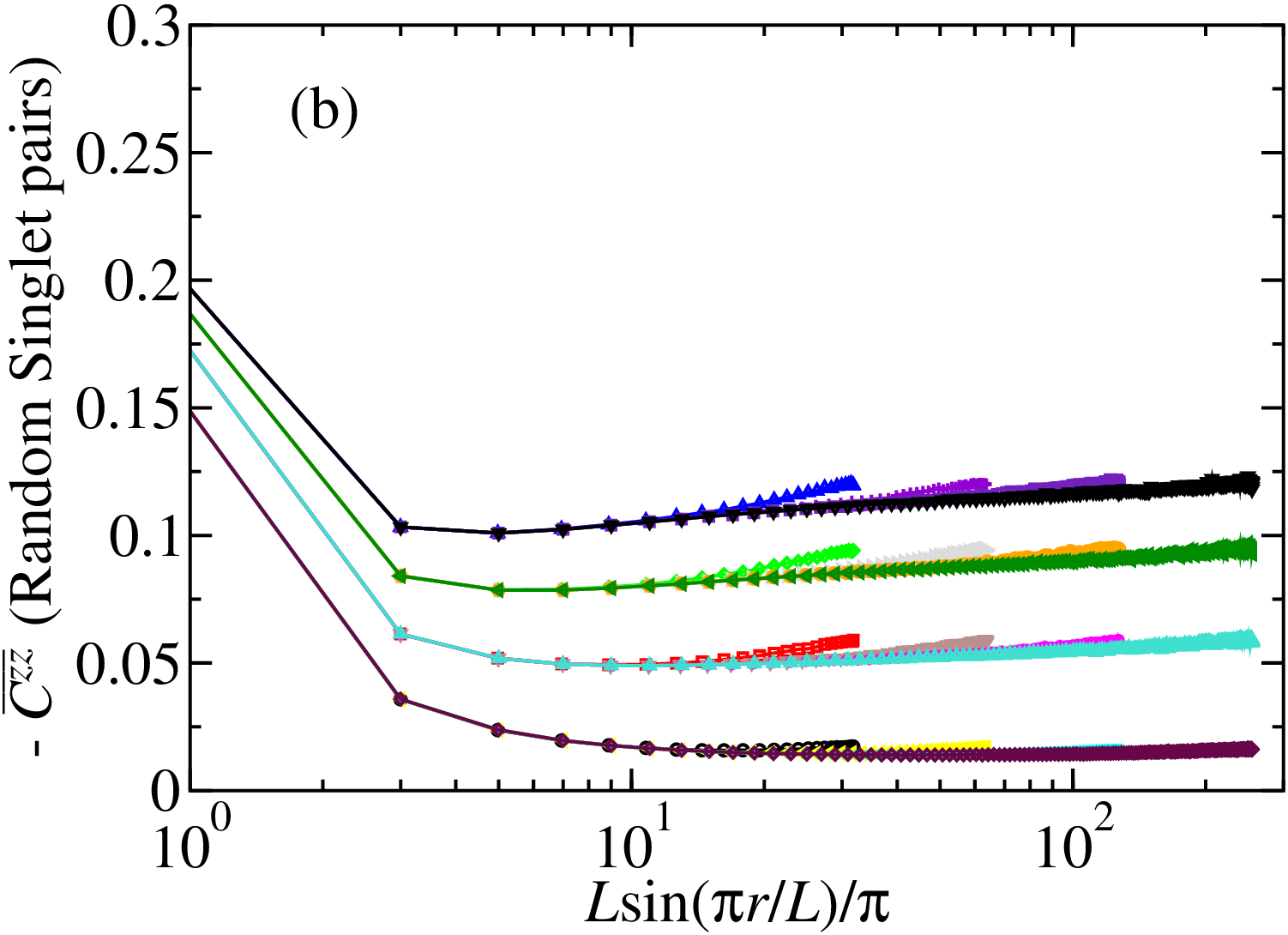}
\par\end{centering}
\caption{\label{fig:RS-correlations}The mean (a) transverse and (b) longitudinal
correlation functions as a function of the chord length $\ell=\frac{L}{\pi}\sin\left(\frac{\pi r}{L}\right)$
considering only the rare singlets of the random-singlet state obtained
via the strong-disorder renormalization-group decimation procedure.
The disorder strengths are $D=0.5$ (bottommost data sets), $1.0$,
$1.5$ and $2.0$ (topmost data sets), and the system sizes are $L=100$
(leftmost data sets), $200$, $400$ and $800$ (rightmost data sets).}
\end{figure}

Finally, and just for completeness, we end our study on the mean correlations
by focusing only on the main culprits for their behavior: the rare
singlet pairs of the random-singlet state (depicted in Fig.~\ref{fig:RS-state}).
Once they are identified (by means of the SDRG decimation procedure~\cite{Fisher1994}),
we compute their mean correlations as a function of the separation
$r$ as shown in Fig.~\ref{fig:RS-correlations}. The naive expectation
based on the clean-disordered crossover is the following. For short
distances $\gamma_{D}r\ll1$ (smaller than the crossover length),
the correlation decays just as in the clean case. For larger distances,
on the other hand, the SDRG singlets become a good approximation and
thus, their correlations are expected to saturate monotonically and
stretched-exponentially fast to a finite value. This expectation is
only partially fulfilled as a non-monotonous saturation is observed.
In addition, the saturation is much slower than one would expect.
(Actually, one could argue that saturation is barely achieved only
for the largest and strongest disordered chains.) 

Finally, we would like to call attention for the importance of using
quadruple precision and having extra care with the numerical instabilities.
Using double precision for $L=800$ and $D=2.0$ yields to data different
from those in Fig.~\ref{fig:RS-correlations}. Surprisingly, the
observed data (not shown here) exhibits a drop in the correlations
(due to the inability of capturing the longest and weakest coupled
spin pairs) compatible with a logarithmic correction of type \eqref{eq:fit-log}
with negative exponent of order $1$, compatible with that reported
in Ref.~\onlinecite{Igloi2000} {[}see Eq.~\eqref{eq:mean-Cx-Igloi}{]}.
Since these rare singlets dominate the mean correlations, this means
that a spurious logarithmic correction can obtained.

\subsection{Typical correlation function and probability distributions\label{sec:typ}}

We now turn our attention to the typical value of the spin-spin correlations
{[}as defined in Eq.~\eqref{eq:typ-C}{]}. In this study, we report
that our data were averaged over $N=10^{5}$ distinct disorder realizations.

We start by assuming that, in the long-distance regime $\gamma_{D}r\gg1$,
the typical correlations can be well approximated by 
\begin{equation}
C_{\text{typ}}^{\alpha\alpha}=\left(-1\right)^{r}c_{\alpha,D}\chi_{\alpha}\left(D,r\right)e^{-A_{\alpha}\sqrt{\gamma_{D}\ell f_{\alpha}\left(r/L\right)}},\label{eq:C-typ-general}
\end{equation}
 where $c_{\alpha,D}$ is a disorder-dependent prefactor, $\chi_{\alpha}$
represents the crossover function (which $\chi_{\alpha}\rightarrow1$
for $\gamma_{D}r\gg1$), $A_{\alpha}$ is a disorder-independent constant,
$\ell$ is the chord length \eqref{eq:chord-length}, $\gamma_{D}$
is the Lyapunov exponent \eqref{eq:Lyapunov-exponent}, and the correction
to the chord length $f_{\alpha}$ is analogous to those for the average
correlations \eqref{eq:f-correction}. Notice that \eqref{eq:C-typ-general}
recovers the SDRG prediction of a stretched exponential decay $\ln\left|C_{\text{typ}}^{\alpha\alpha}\right|\sim-r^{\psi_{\alpha}}$
with universal (disorder-independent) and isotropic exponent $\psi_{\alpha}=\psi=\frac{1}{2}$.
Equation \eqref{eq:C-typ-general} assumes that disorder enters in
the exponential only via the Lyapunov exponent $\gamma_{D}$. While
its presence is natural since the stretched exponential form requires
a length scale, and thus the corresponding Lyapunov exponent of the
underlying single-parameter scaling theory~\cite{Hoyos2014}, it
is not clear why $A_{\alpha}$ and $f_{\alpha}$ should be disorder
independent. Nonetheless, as we show below, this hypothesis is compatible
with our data. Finally, we mention that, unlike the mean correlations,
the numerical prefactor $c_{x,D}$ is the same for both even and odd
separations $r$.

\begin{figure}[t]
\begin{centering}
\includegraphics[clip,width=0.85\columnwidth]{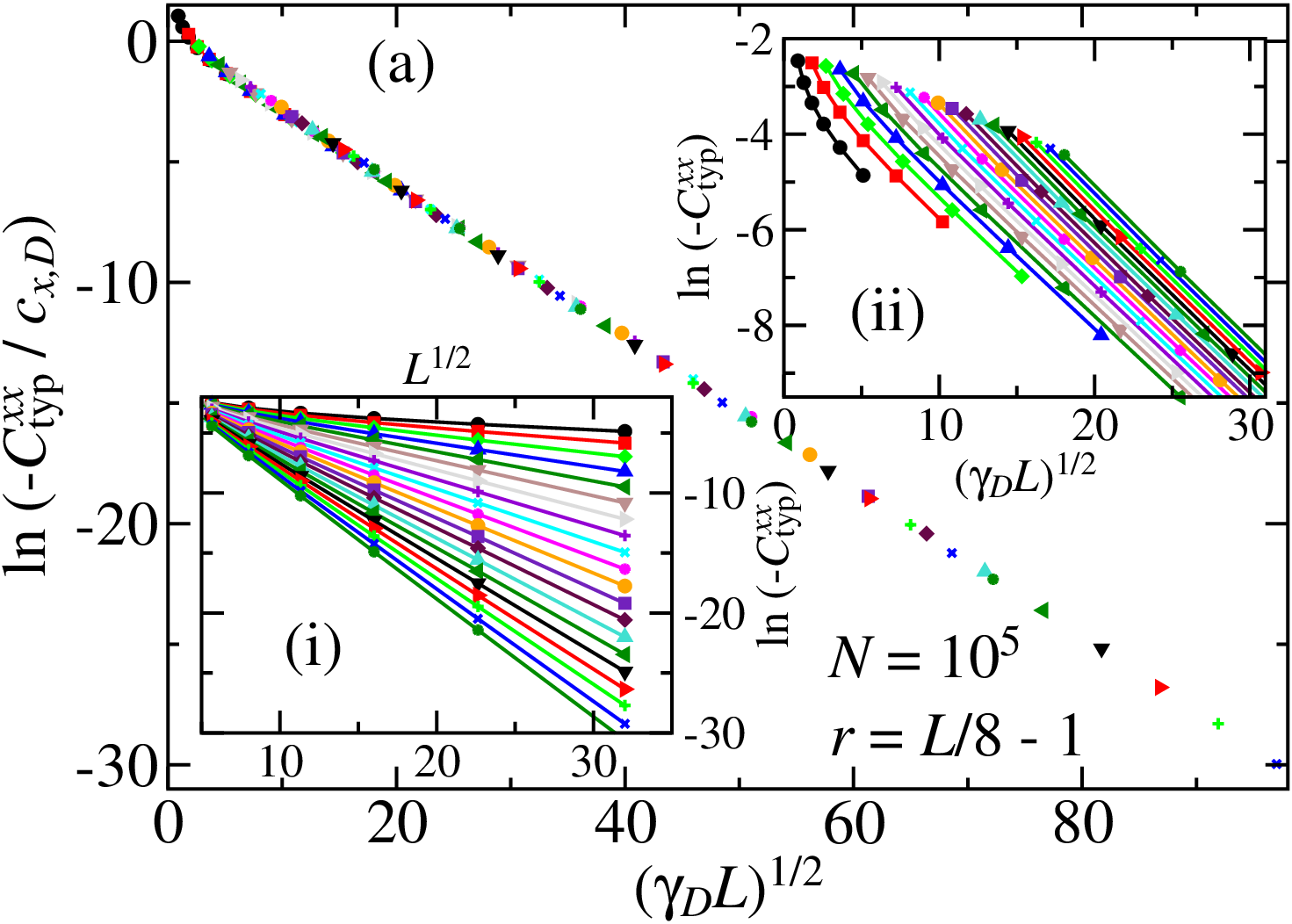}\\
\includegraphics[clip,width=0.85\columnwidth]{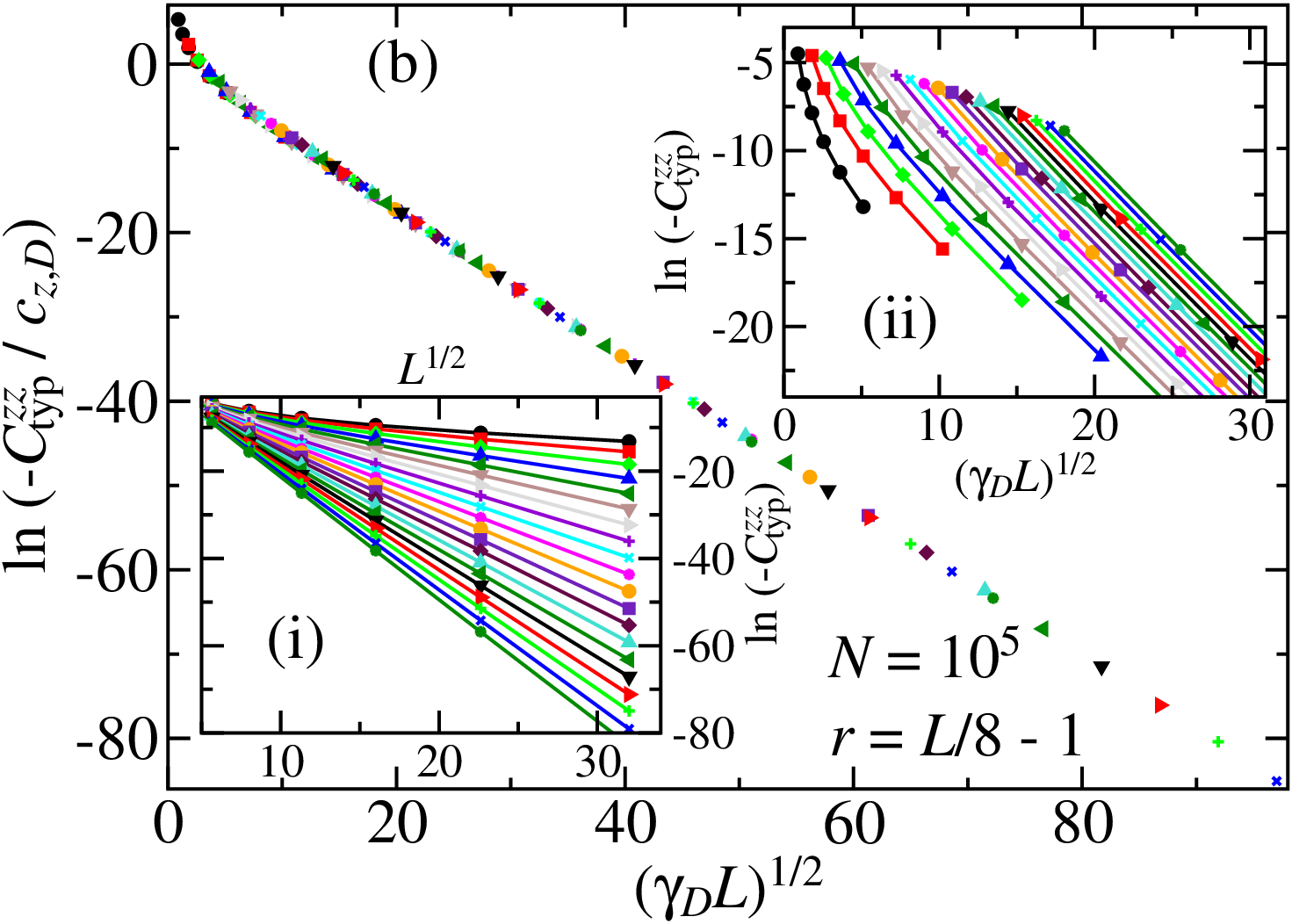}\\
\includegraphics[clip,width=0.8\columnwidth]{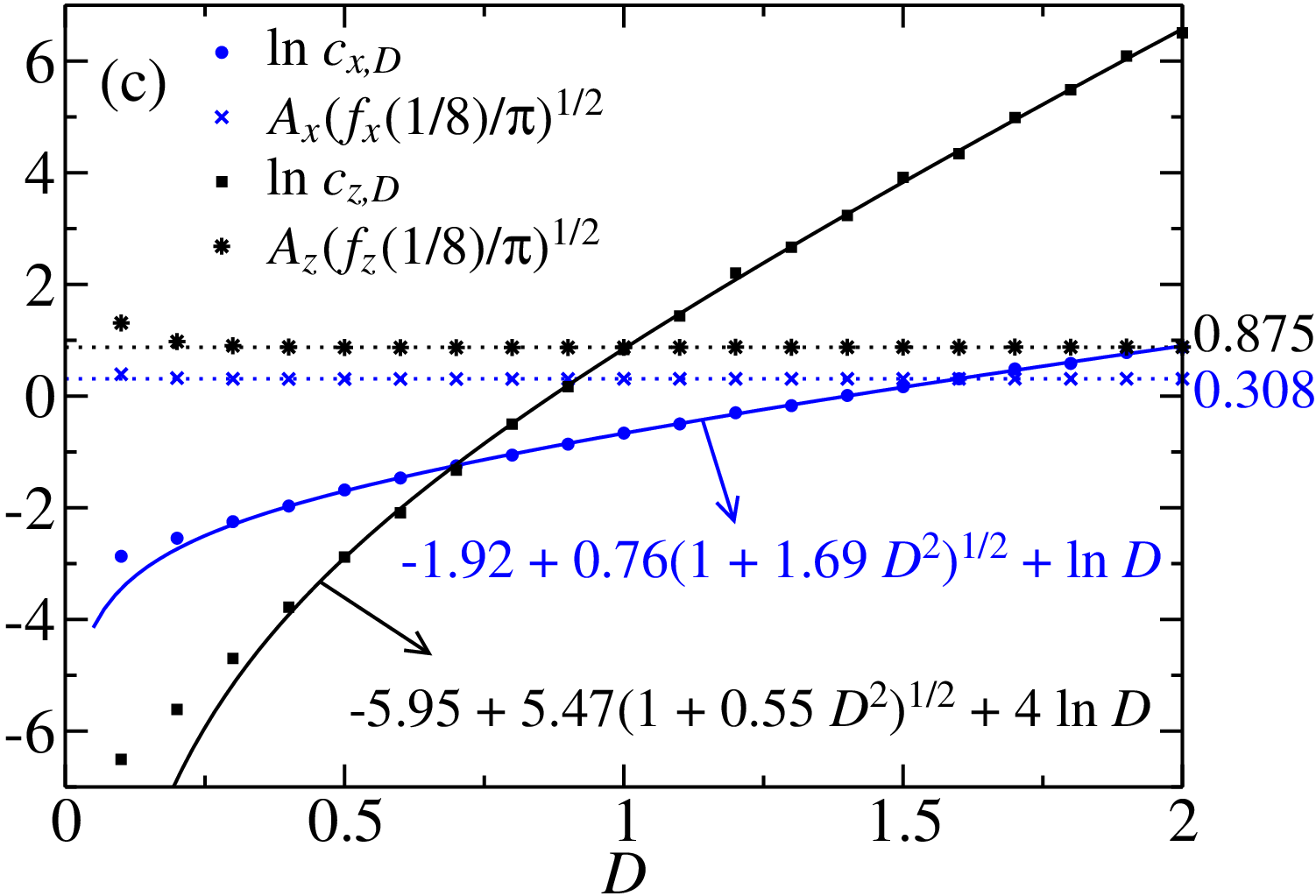}
\par\end{centering}
\caption{\label{fig:typ2}Uncorrelated disorder model: the typical value of
the (a) transverse and (b) longitudinal spin-spin correlations as
a function of the system size $L$ {[}inset (i){]} and $\gamma_{D}L$
{[}close up in inset (ii){]}, for various disorder strengths $D$
and separation $r=L/8-1$. System sizes are $L=2^{n}$, with $n$
ranging from $5$ to $10$. Disorder strength varies from $D=0.1$
to $2.0$ {[}topmost and bottommost curves in inset (i), respectively{]}
in equal steps of $0.1$. The lines are guide to the eyes. In panel
(c), $c_{\alpha,D}$ and $A_{\alpha}$ are plotted against $D$. They
are the best fits \eqref{eq:C-typ-general} to the data in insets
(ii) (restricted to $\left|C_{\text{typ}}^{xx}\right|<2.5\,10^{-2}$
and $\left|C_{\text{typ}}^{zz}\right|<2.0\,10^{-4}$). The solid lines
are the best fits to Eq.~\eqref{eq:prefactor-c} restricted to $D\protect\geq0.4$
and are used to obtain the data collapse in the main plots of panels
(a) and (b).}
\end{figure}

In Figs.~\hyperref[fig:typ2]{\ref{fig:typ2}(a)} and \hyperref[fig:typ2]{(b)}
we plot respectively the transverse and longitudinal typical correlations
for $r=L/8-1$ and various chain sizes $L$ and disorder strengths
$D$. The insets (i) of those figures bring the raw data from which
the SDRG prediction $\ln\left|C_{\text{typ}}^{\alpha\alpha}\right|\sim-\sqrt{r}$
is confirmed. 

We then replot the correlations as a function of $\left(\gamma_{D}L\right)^{\psi}$
as shown in the insets (ii) of those figures. Apparently, the constant
$A_{\alpha}$ is disorder-independent. Moreover, the values of $L$
used seem to be sufficiently large (at least for $D\geq0.4$) such
that $\chi_{\alpha,D}\approx1$. Therefore, it is safe to obtain the
values of $c_{\alpha,D}$ and $A_{\alpha}$ by simply fitting Eq.~\eqref{eq:C-typ-general}
to our data.~\footnote{We consider only the data such that $\left|C_{\text{typ}}^{xx}\right|<2.5\,10^{-2}$
and $\left|C_{\text{typ}}^{zz}\right|<2.0\,10^{-4}$. This is simply
to ensure some meaning to the fitting function \eqref{eq:C-typ-general}
when disorder is weak ($D<0.6$). As we explain latter on, this has
no influence in our results.} The fitting values of $c_{\alpha,D}$ and $A_{\alpha}\sqrt{f_{\alpha}\left(1/8\right)}$
are plotted in Fig.~\hyperref[fig:typ2]{\ref{fig:typ2}(c)}. For
$D\leq0.3$, these are effective values {[}not in the asymptotic regime
$\gamma_{D}r\gg1$, as can be seen in insets (ii){]}. The fitting
values are consistent with $A_{\alpha}$ and $f_{\alpha}$ being disorder
independent. 

In order to proceed, as in the analysis of the mean correlation $\overline{C^{xx}}$,
we need the relation between the numerical prefactor $c_{\alpha,D}$
and the disorder strength $D$. Clearly, one needs a theory capable
of capturing both the clean and the disorder critical behaviors. Here,
however, we will simply try to connect the clean behavior $C_{\text{clean}}^{\alpha\alpha}\sim c_{1,\alpha}r^{-\eta_{\alpha}}$
(with $\eta_{x}=1/2$ and $\eta_{z}=2$) to the strong-disorder one
$C_{\text{typ}}^{\alpha\alpha}\sim c_{\alpha,D}e^{-c_{2,\alpha}\sqrt{\gamma_{D}r}}$.
Assuming a sharp crossover at $r=r_{\alpha}^{*}=c_{3,\alpha}\gamma_{D}^{-1}$,
continuity requires that $\ln c_{\alpha,D}=p_{\alpha}+2\phi_{\alpha}\ln D$.
However, this poorly fits the data in Fig.~\hyperref[fig:typ2]{\ref{fig:typ2}(c)}.
We have tried several modifications of this scenario in order to improve
the fit. They include adding one or two polynomials $\propto D^{n}$
and power-laws $\propto D^{-n}$, and also changing the prefactor
of the logarithmic term. The worst modifications are those in which
the logarithmic term is dropped out, implying that $c_{\alpha,D}\propto D^{2\phi_{\alpha}}$
is very robust. The most successful modification is such that we admit
a sharp crossover happening at $r_{\alpha}^{*}=c_{3,\alpha}\gamma_{D}^{-1}+c_{4,\alpha}$.
The exponential in the typical correlation then acquires a dependence
on $D$ yielding to 
\begin{equation}
\ln c_{\alpha,D}=o_{\alpha}+p_{\alpha}\sqrt{1+q_{\alpha}D^{2}}+2\phi_{\alpha}\ln D,\label{eq:prefactor-c}
\end{equation}
 with $o_{\alpha}$, $p_{\alpha}$ and $q_{\alpha}$ being fitting
parameters. The fitting values are shown in Fig.~\hyperref[fig:typ2]{\ref{fig:typ2}(c)}
for which only the data for $D\geq0.4$ were used. The reason is that
for smaller values of $D$, the slope $A_{\alpha}$ is not fully saturated
(due to the effects of the crossover function $\chi_{\alpha}$). We
have checked that changing any of the fitting parameters values by
$5\%$ does not change the quality of the fit, i.e., the reduced weighted
error sum $\bar{\chi}^{2}$remains the same within the statistical
error. This means that $5\%$ is a reasonable estimate for the accuracy
of our fit.

We now put Eq.~\eqref{eq:prefactor-c} to test by assuming that it
holds for all disorder strengths. In the main panels of Figs.~\hyperref[fig:typ2]{\ref{fig:typ2}(a)}
and \hyperref[fig:typ2]{(b)}, we plot $C_{\text{typ}}^{\alpha\alpha}/c_{\alpha,D}$
as a function of $\sqrt{\gamma_{D}L}$ (recall $r/L\approx1/8$ is
fixed). Remarkably, and somewhat surprisingly, we obtain good data
collapse even for the least disordered system studied $D=0.1$. For
small $\sqrt{\gamma_{D}L}$, $C_{\text{typ}}^{\alpha\alpha}/c_{\alpha,D}$
deviates from the pure stretched exponential, which is attributed
to the crossover term $\chi_{\alpha,D}\left(r\right)$. The fact that
the data collapse for all disorder strengths means that, to a good
approximation, disorder enters in $\chi_{\alpha,D}$ through the combination
$\gamma_{D}r$, i.e., $\chi_{\alpha,D}(r)=\chi_{\alpha}\left(\gamma_{D}r\right)$.
Once more, this data collapse also supports that $f_{\alpha}$ is
a disorder-independent function.

\begin{figure}[t]
\begin{centering}
\includegraphics[clip,width=0.8\columnwidth]{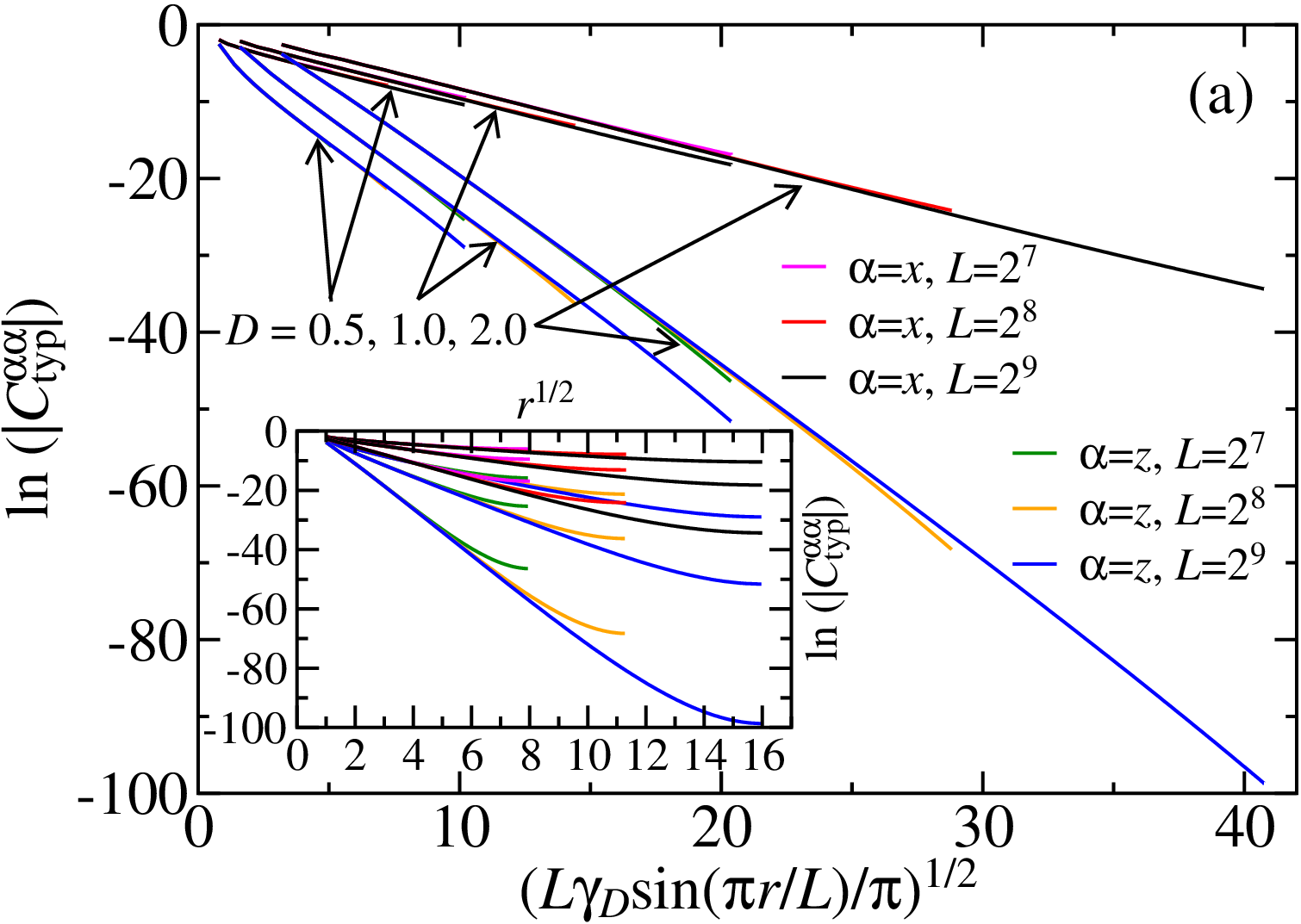}\\
\includegraphics[clip,width=0.8\columnwidth]{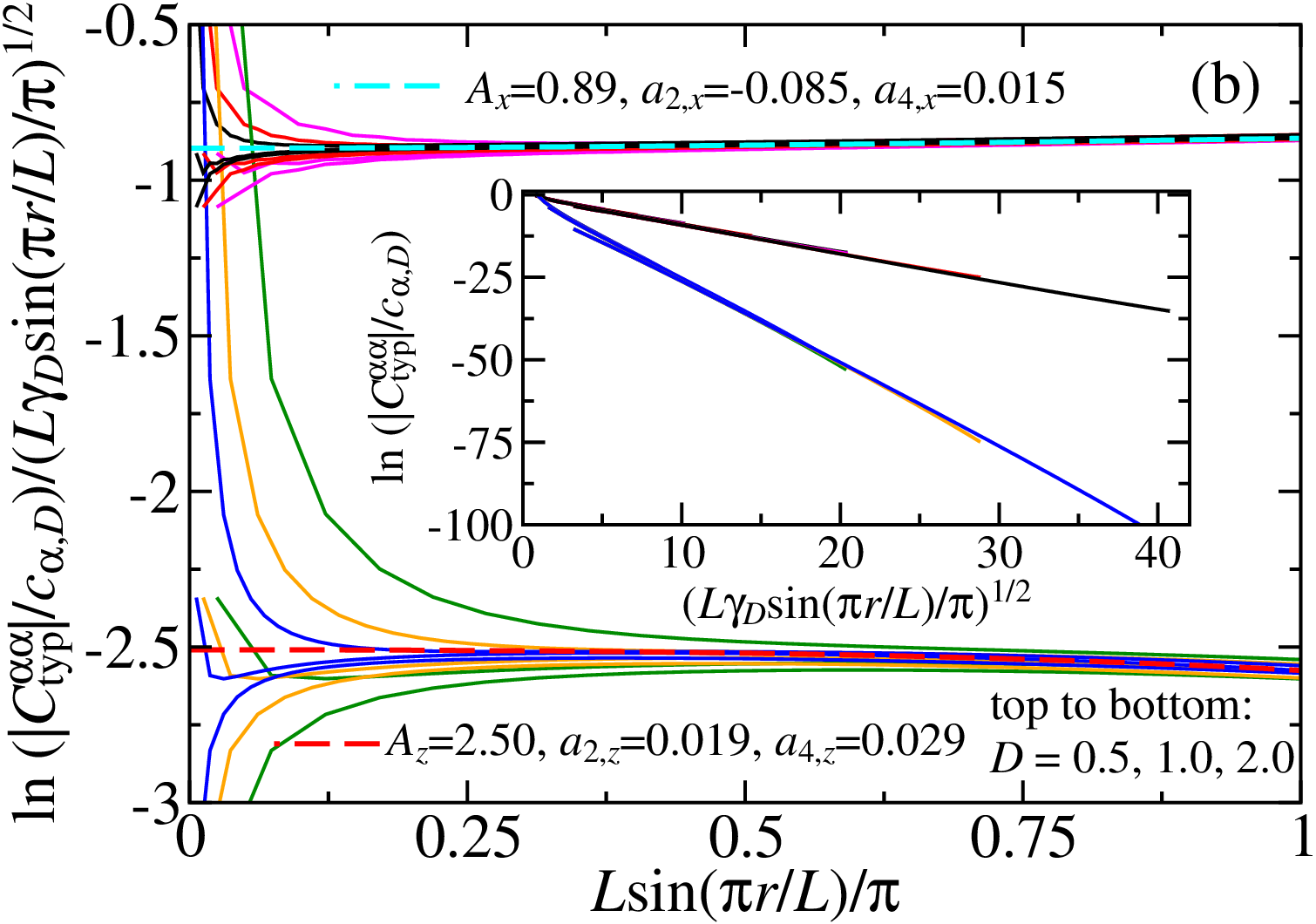}
\par\end{centering}
\caption{\label{fig:typ} The transverse and longitudinal typical value of
the correlation function $C_{{\rm typ}}^{\alpha\alpha}(r)$ for chain
sizes $L=2^{n}$, with $n=7,\ 8,$ and $9$, and disorder strengths
$D=0.5,\ 1.0,$ and $D=2.0$. The corresponding legends are given
in panel (a). The same data were plotted in different ways in the
main panels and insets (see text). The dashed lines in panel (b) are
the best fits according to Eq.~\eqref{eq:C-typ-general}. }
\end{figure}

We are now in the position to study the chord-length-correction function
$f_{\alpha}$ in Eq.~\eqref{eq:C-typ-general}. In Fig.~\ref{fig:typ},
we plot $C_{\text{typ}}^{\alpha\alpha}$ as a function of suitable
combinations of $r$, $L$ and $\gamma_{D}$. For clarity, we show
only a few data such as $L=2^{n}$, with $n=7,$ $8$, and $9$, and
$D=0.5$, $1.0$, and $2.0$. As can be seen in the main panel of
Fig.~\hyperref[fig:typ]{\ref{fig:typ}(a)}, the chord length $\ell=\frac{L}{\pi}\sin\left(\pi r/L\right)$
is nearly enough for accounting all the finite-size effects. The combination
$\gamma_{D}\ell$ {[}see the inset of Fig.~\hyperref[fig:typ]{\ref{fig:typ}(b)}{]}
nearly collapses all the data. The role played by the crossover function
$\chi_{\alpha,D}$ and the chord-length-correction function $f_{\alpha,D}$
are shown in the main panel of Fig.~\hyperref[fig:typ]{\ref{fig:typ}(b)}.
We note that all curves converge to a single one in the large-$L$
limit, in agreement with the hypothesis \eqref{eq:C-typ-general}.
The dashed lines are the best fits restricted to the region $\ell=\frac{L}{\pi}\sin\left(\frac{\pi r}{L}\right)>0.5$
and considering only the large system size $L=512$. The fitting values
of $A_{\alpha}$ and $a_{2n,\alpha}$ are reported in Fig.~\hyperref[fig:typ]{\ref{fig:typ}(b)}.
(Adding higher order terms do not improve our fit.) Notice the small
correction to the chord length $a_{2,4,\alpha}\ll1$, much smaller
than those for the mean correlations. Interestingly, the crossover
function $\chi_{\alpha,D}$ is non-monotonic with respect to $D$.
Notice that $\chi_{\alpha,D}$ tends to $1$ from above for weak disorder
$D\lessapprox1.0$, and from below otherwise. Finally, we verified
(not shown) that $\left|\chi_{x,D}-1\right|/\left|\chi_{z,D}-1\right|$
is nearly a constant for large $\gamma_{D}r$.

\begin{figure}[b]
\begin{centering}
\includegraphics[clip,width=0.8\columnwidth]{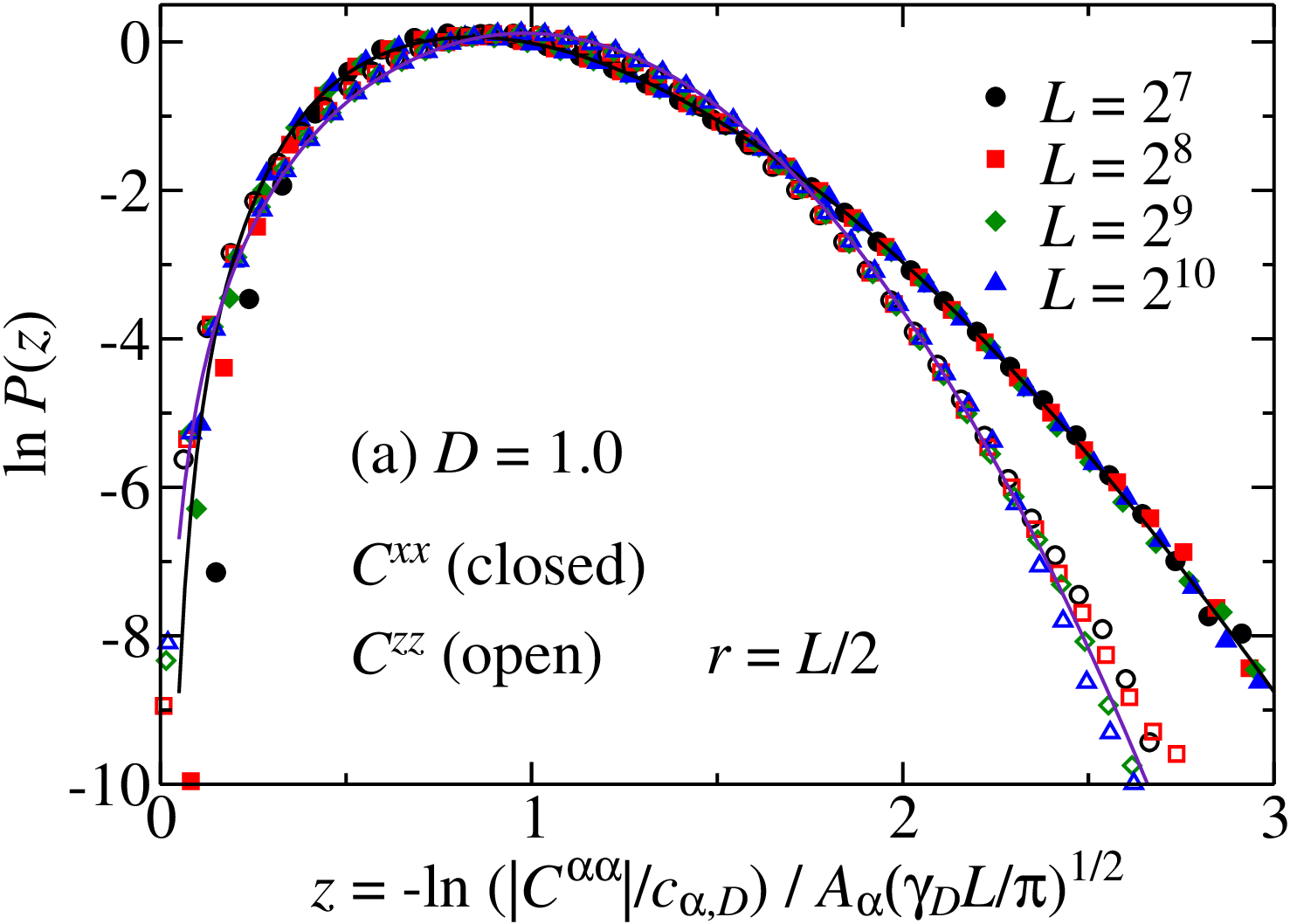}\\
\includegraphics[clip,width=0.8\columnwidth]{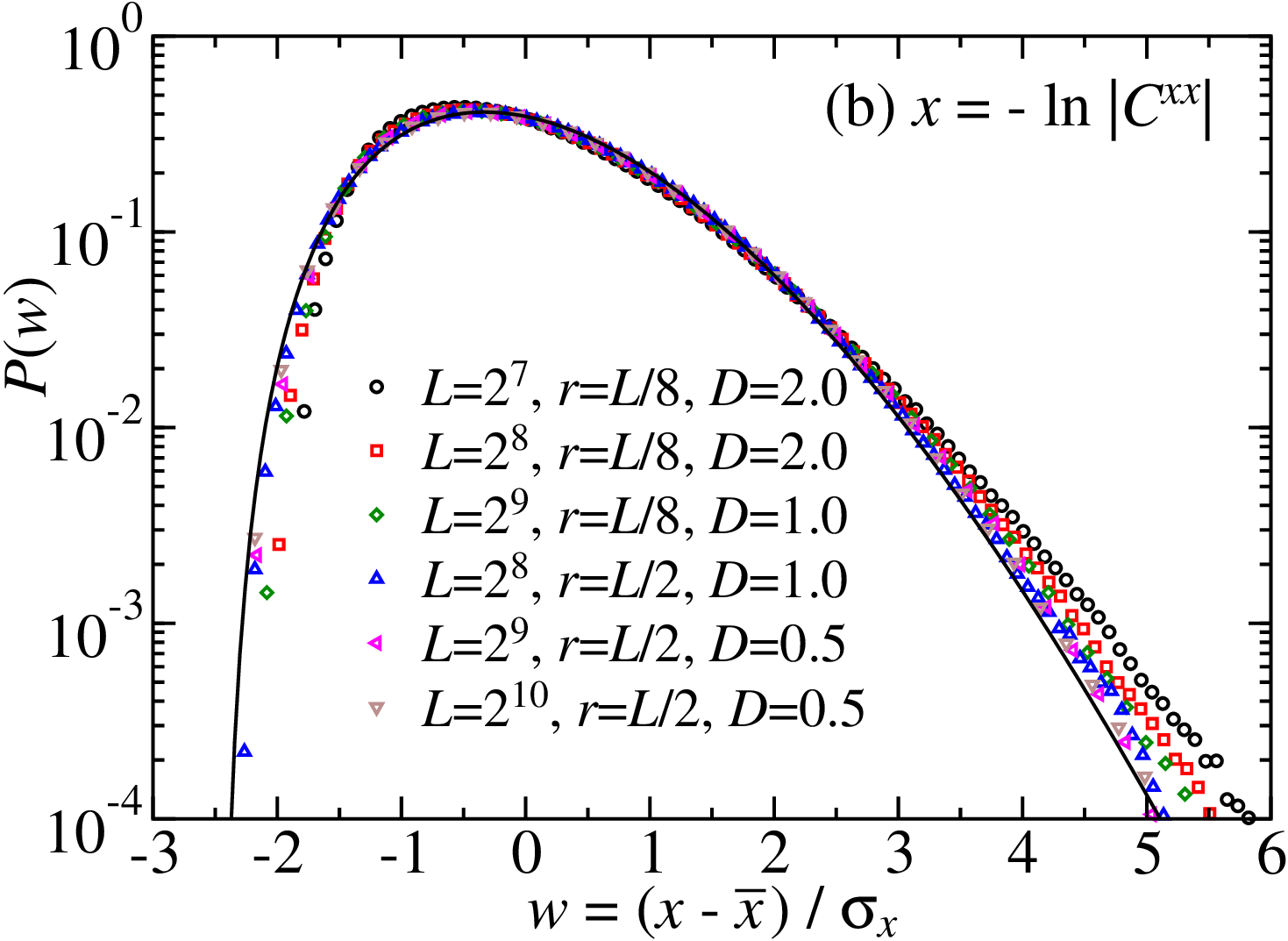}\\
\includegraphics[clip,width=0.8\columnwidth]{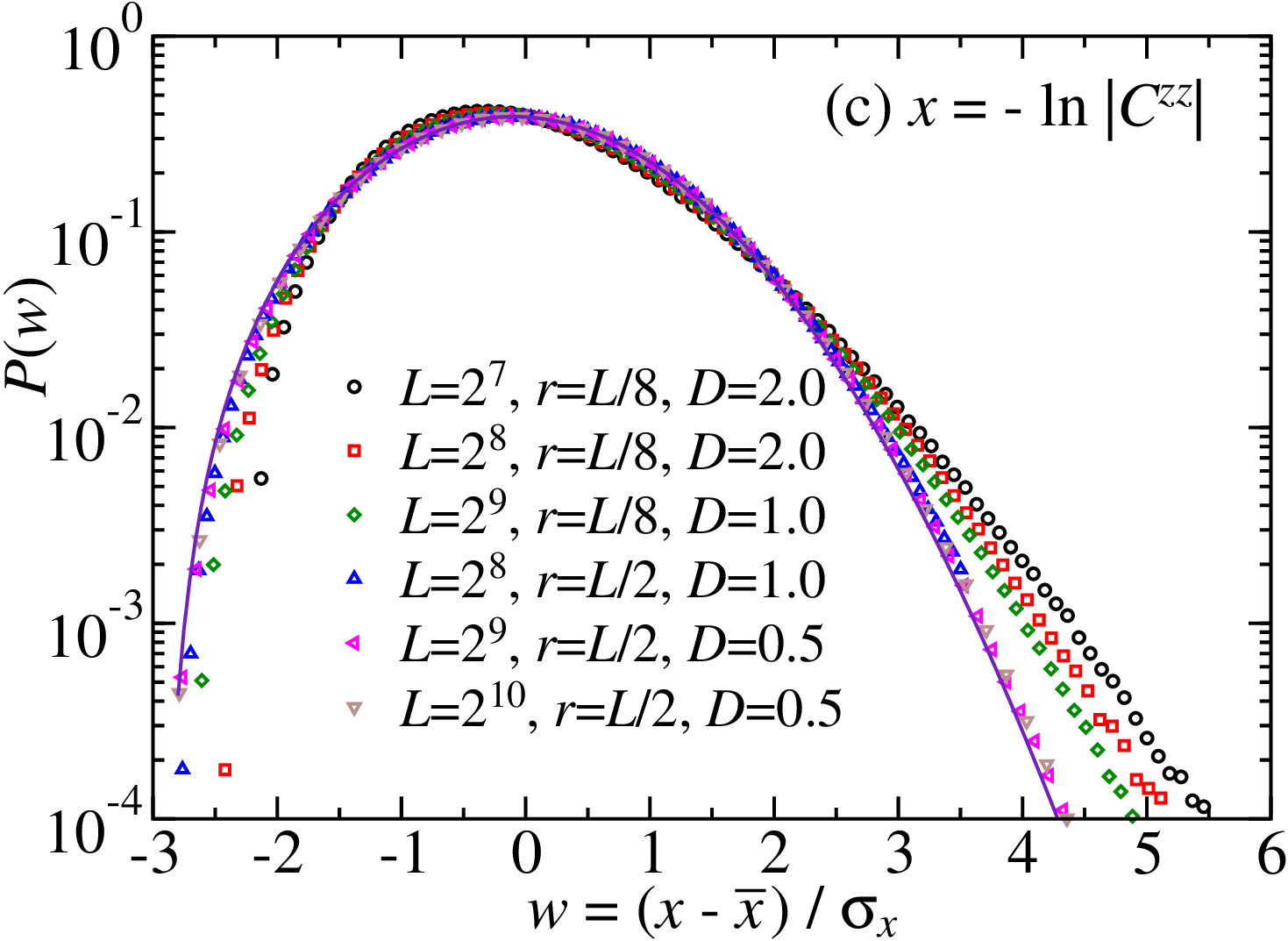}
\par\end{centering}
\caption{\label{fig:dist}Rescaled distribution of the correlation function
$C^{\alpha\alpha}(r)$, for (a) fixed distance $r=L/2$ and disorder
strength $D=1.0$, and (b) and (c) distances $r=L/8$ and $L/2$ and
various disorder strengths $D$ for various system sizes $L$. The
lines are best fits of Eq.~\eqref{eq:dist-P} (see text for details).}
\end{figure}

We now turn our attention to the correlation function distribution.
In the pioneering work by Fisher, it was conjectured that $\ln\left|C^{\alpha\alpha}\right|/\sqrt{r}$
converges to a non-trivial distribution for large separation $r$.
This conjecture was confirmed in Refs.~\onlinecite{Young1996,Henelius1998}
by numerically computing the distribution of $\ln\left|C^{\alpha\alpha}\right|/\sqrt{r}$
for a fixed disorder strength $D$. We confirm this conjecture by
studying the distribution of $z=-\ln\left(\left|C^{\alpha\alpha}\right|/c_{\alpha,D}\right)/A_{\alpha}\sqrt{\gamma_{D}L/\pi}$
for fixed separation $r=L/2$ and disorder strength $D=1.0$, for
various system sizes $L$. As shown in Fig.~\hyperref[fig:dist]{\ref{fig:dist}(a)}
the distribution $P_{\alpha}\left(z\right)$ converges to a non-trivial
one for large $L$. According to Eq.~\eqref{eq:C-typ-general}, the
first moment of $P_{x}$ and $P_{z}$ converges to the unity in the
$\gamma_{D}r\gg1$ regime. We first notice that $P_{x}$ and $P_{z}$
are not equal. Also, both distributions are narrow. We have tried
many different fitting functions. Since they are narrow, we tried
Weibull and Gaussian distributions but with poor success. The most
satisfactory one is 
\begin{equation}
P_{\alpha}\left(z\right)=C_{\alpha}\exp\left(-\left|\frac{z}{\delta_{\alpha}}-\zeta_{\alpha}\right|^{\gamma_{\alpha}}+b_{\alpha}\left(\frac{\delta_{\alpha}}{z}\right)^{\gamma_{\alpha}^{\prime}}\right),\label{eq:dist-P}
\end{equation}
 where $C_{\alpha}$ is a normalization constant, and $\delta_{\alpha},$
$\zeta_{\alpha}$, $b_{\alpha}$, $\gamma_{\alpha}$ and $\gamma_{\alpha}^{\prime}$
are fitting parameters with obvious interpretations. The first term
in the exponential dictates the large-distance behavior $\gamma_{D}r\gg1$
which, naively, we expect to be near a Gaussian. Then, $\gamma_{\alpha}$
is the corresponding exponent for the tail, $\delta_{\alpha}$ would
represent the width and $\zeta_{\alpha}$ the rescaled offset. The
second term dictates the low-distance behavior $\gamma_{D}r\ll1$
with corresponding exponent $\gamma_{\alpha}^{\prime}$.~\footnote{We also have tried polynomials $P_{\alpha}\propto z^{\lambda_{\alpha}^{\prime}}$
and verified satisfactory fits with $\lambda_{x}^{\prime}\approx5\pm1$
and $\lambda_{z}^{\prime}\approx2.6\pm0.6$.} Notice that this term represents a sharp cutoff for $z<0$. We have
tried to offset this term via $z-z_{0}$ and found that $\left|z_{0}\right|\lessapprox0.02$.
Surprisingly, our choice of $z$ makes the $P_{\alpha}\left(z<0\right)=0$.
Our fits extrapolated to $L\rightarrow\infty$ are shown as solid
lines in Fig.~\hyperref[fig:dist]{\ref{fig:dist}(a)} and are numerically
equal to $\zeta_{x}=0.65(5)$, $\zeta_{z}=1.1(2)$, $\delta_{x}=0.64(3)$,
$\delta_{z}=0.62(2)$, $\gamma_{x}=1.71(3)$, $\gamma_{z}=2.11(5)$,
$b_{x}=4.6(3)$, $b_{z}=8.7(5)$, $\gamma_{x}^{\prime}=0.41(3)$,
and $\gamma_{z}^{\prime}=0.19(2)$.

We now step forward and study how $P_{\alpha}$ depends on $D$. In
Figs.~\hyperref[fig:dist]{\ref{fig:dist}(b)} and \hyperref[fig:dist]{(c)}
we plot the distribution of $-\ln\left|C^{\alpha\alpha}\right|$ (shifted
by its average and divided by its standard deviation) for various
disorder strengths $D$, system sizes $L$, and separations $r=L/8$
and $L/2$. For comparison, we replot the corresponding fits of panel
\hyperref[fig:dist]{(a)} in panels \hyperref[fig:dist]{(b)} and
\hyperref[fig:dist]{(c)}, also shifted by the corresponding mean
values (which are both equal to one) and divided by the corresponding
standard deviation ($0.38$ and $0.35$, respectively for $\alpha=x$
and $\alpha=z$). For separations $r=L/2$, all distributions are
clearly universal, i.e., disorder independent. For shorter separations
$r=L/8$, the distributions differ from the universal one. Given the
systematic tendency towards the universal distribution for larger
and larger system sizes $L$, we then attribute this discrepancy to
the fact that the limit of large separation has not been achieved
for those cases. We therefore conclude that, in the large separation
limit, the distribution of $\ln\left|C^{\alpha\alpha}\right|/\sqrt{\gamma_{D}r}$
converges to a non-trivial, narrow and universal (disorder-independent)
distribution.

\section{Spin-spin correlations for the correlated coupling constants model\label{sec:corr}}

In this section we report our results on the average and typical correlation
functions for the case of correlated disorder in the model \eqref{eq:H},
which is defined by a set of coupling constants $\{J_{1},J_{1},J_{2},J_{2},\dots,J_{L/2},J_{L/2}\}$
(see Sec.~\ref{sec:methods}). 

\begin{figure}[b]
\begin{centering}
\includegraphics[clip,width=0.8\columnwidth]{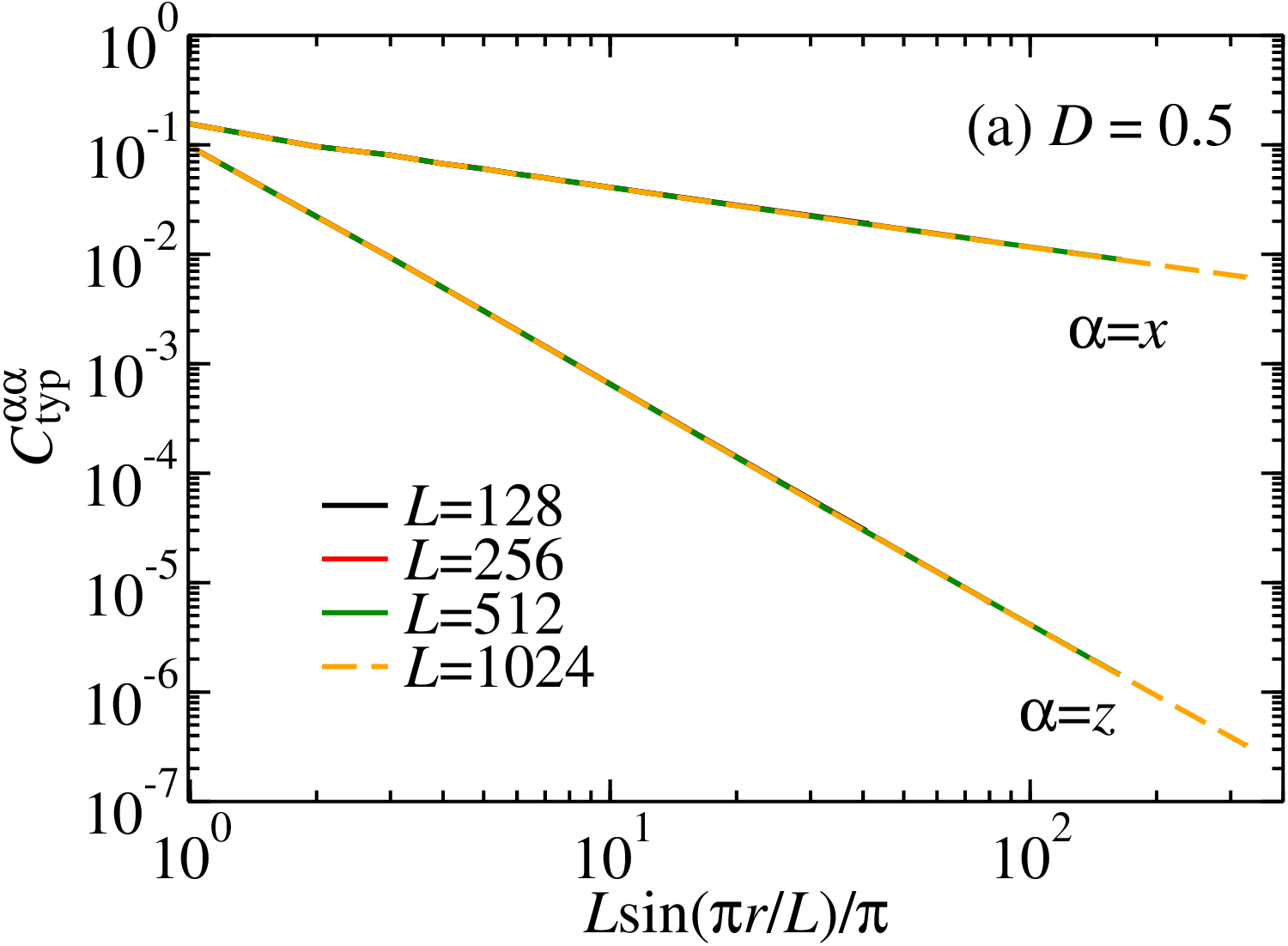}\\
\includegraphics[clip,width=0.8\columnwidth]{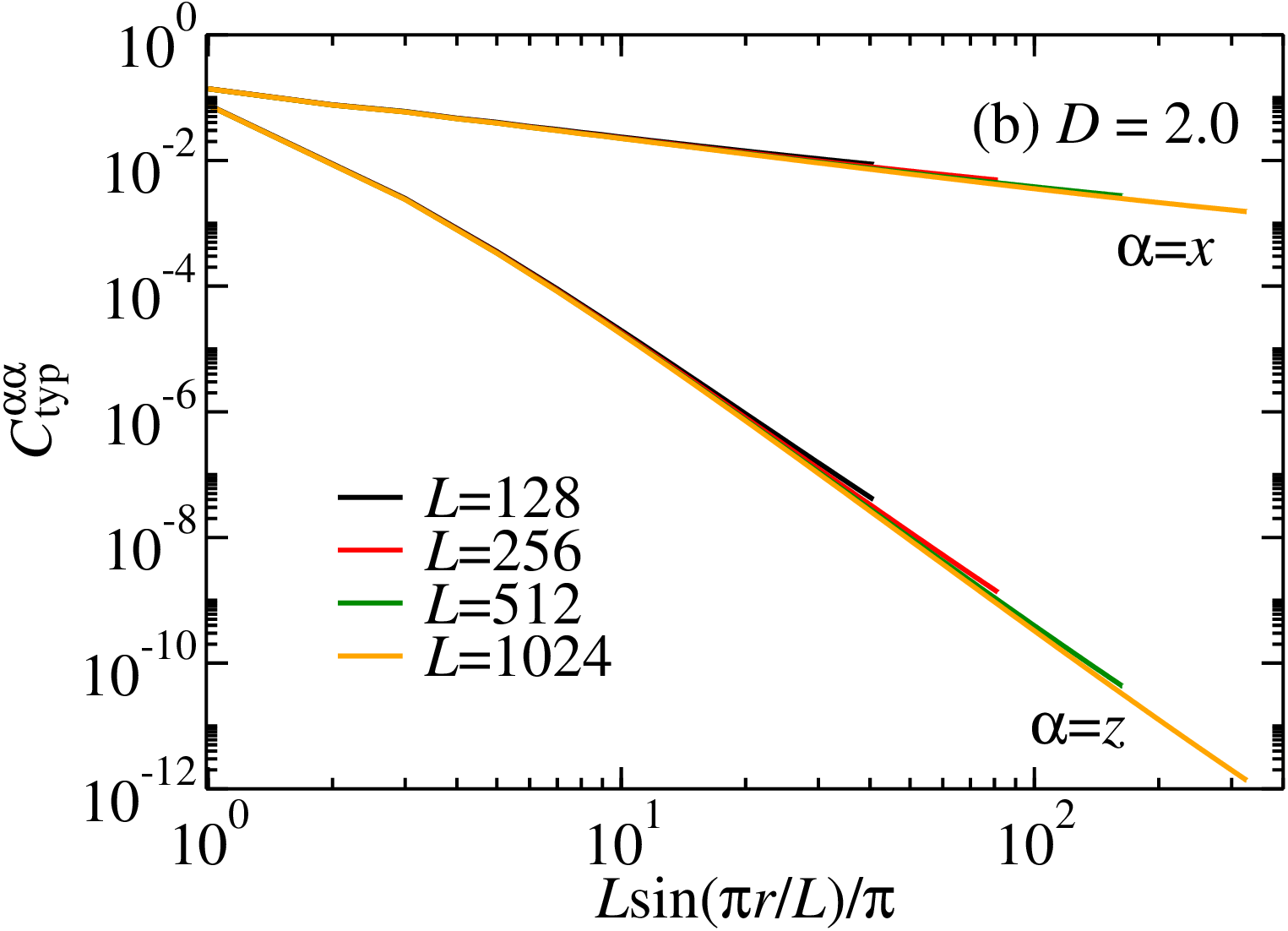}
\par\end{centering}
\caption{\label{fig:ctyp} The typical correlation functions as a function
of the chord length \eqref{eq:chord-length} for the case of correlated
disorder. The data were averaged over $N=10^{5}$ disorder realizations.}
\end{figure}

Unlike the uncorrelated disorder model, there is no analytical theory
predicting the critical exponents. Here, our purpose is to determine
them for the average and typical correlation functions.

In Fig.~\ref{fig:ctyp}, we plot the typical correlations for various
chain lengths $L$ and disorder strengths $D=0.5$ and $1.0$. Clearly,
the chord length \ref{eq:chord-length} is almost a perfect scaling
variable.~\footnote{Likewise, the chord length was verified to be nearly the correct scaling
variable for the Rényi entanglement entropy for any disorder strength
$D$~\cite{Getelina2016}.} 

\begin{figure}[t]
\begin{centering}
\includegraphics[clip,width=0.8\columnwidth]{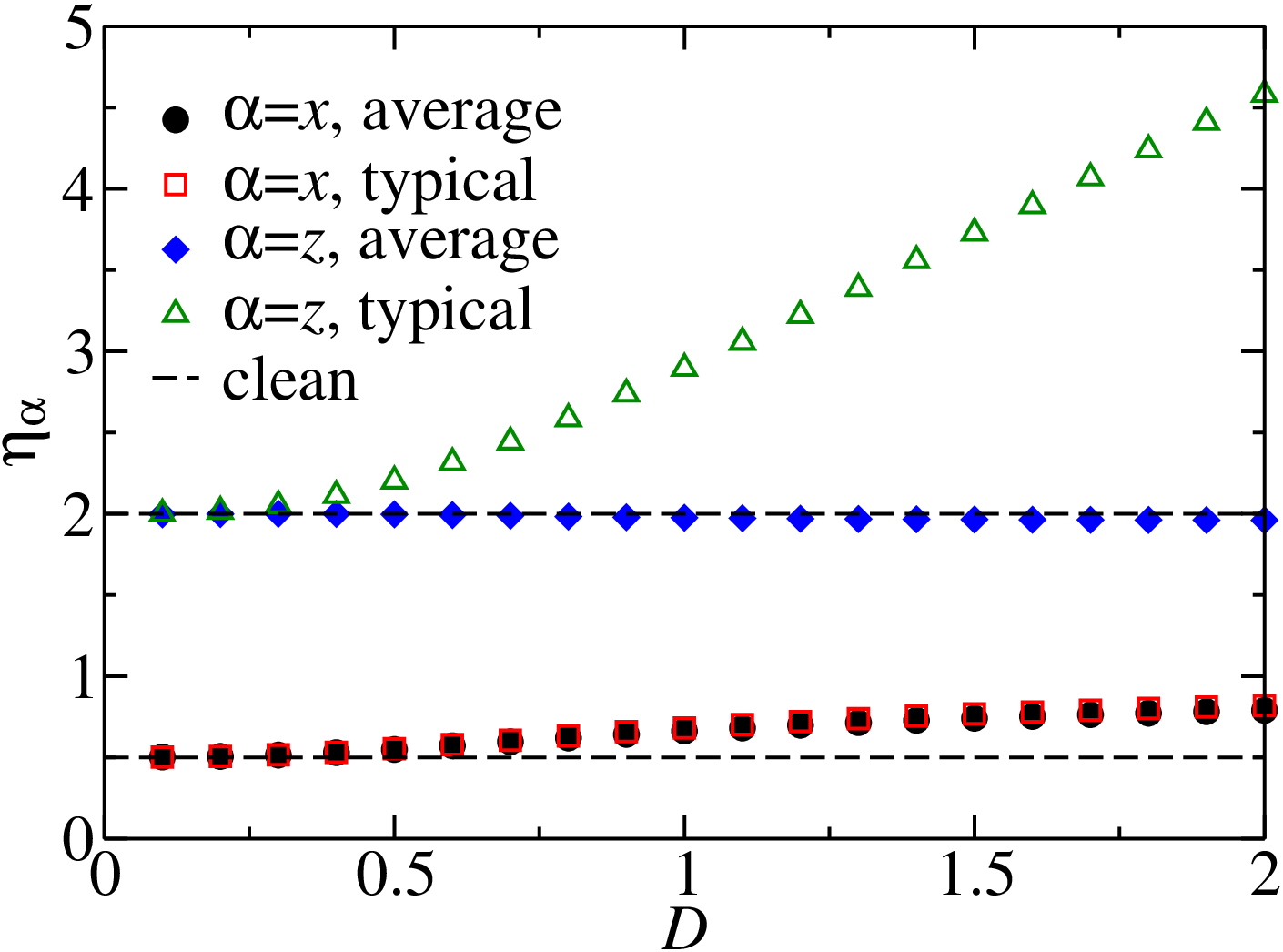} 
\par\end{centering}
\caption{\label{fig:exponent}The average and typical correlation function
critical exponents $\eta_{x,z}$ as a function of the disorder strength
$D$ for the correlated disorder model. The dashed lines correspond
to the homogeneous (clean) system values.}
\end{figure}

Clearly, the typical correlations decay algebraically, which is very
distinct from their uncorrelated disorder counterpart. Evidently (not
shown here), the average correlations also decays algebraically with
the spin separation $r$. Simple fits restricted to the long-distance
tail provide the corresponding exponents which are shown in Fig.~\ref{fig:exponent}. 

For disorder strengths below the threshold $D_{c}\approx0.3$, the
exponent agrees with those of the clean system $\eta_{z}=4\eta_{x}=2$,
as expected. Tuning the line of finite-disorder fixed points by increasing
$D$ beyond $D_{c}$, the exponents vary continuously and in a non-trivial
fashion. 

\begin{figure}[b]
\begin{centering}
\includegraphics[clip,width=0.8\columnwidth]{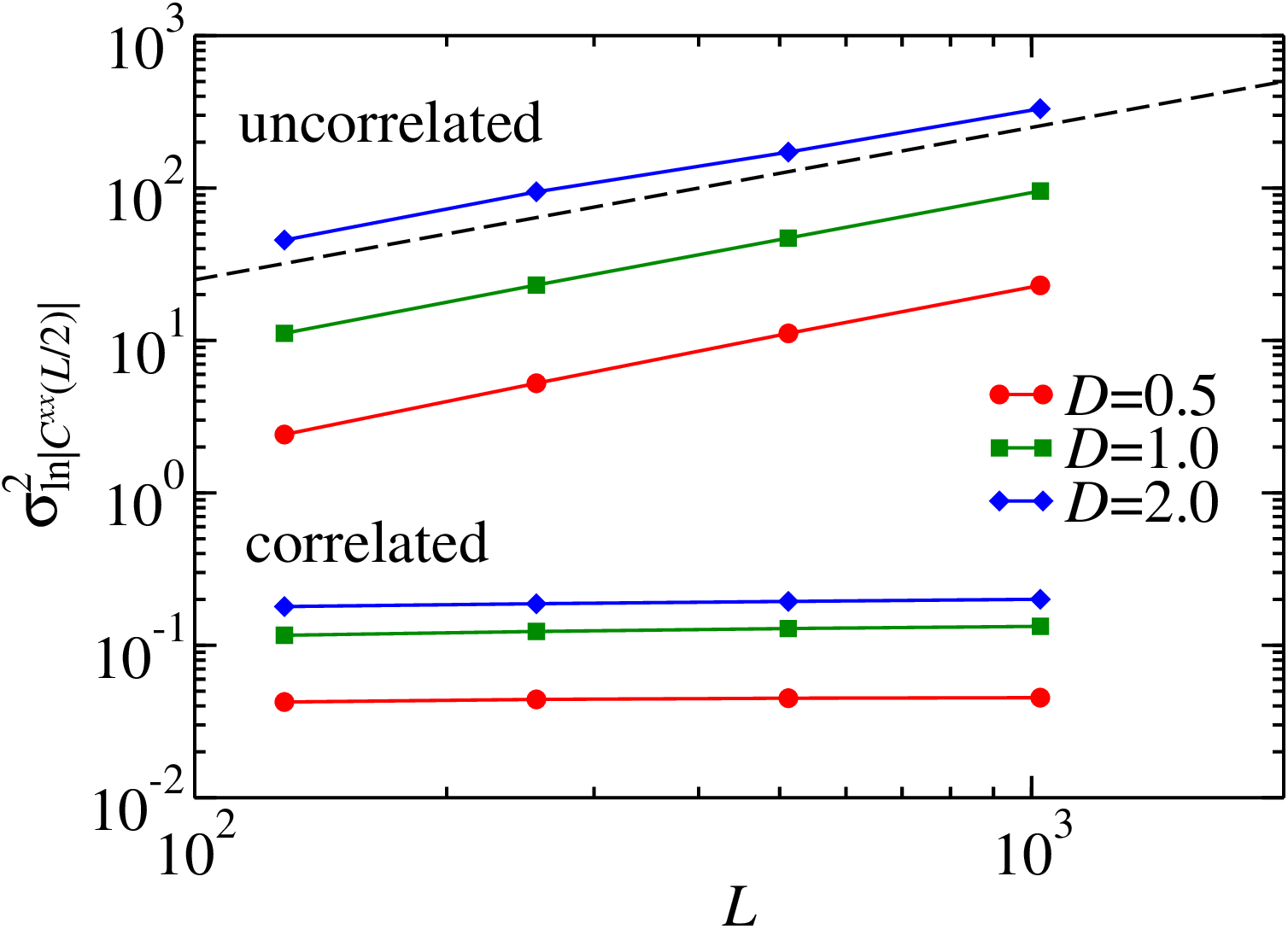}
\par\end{centering}
\caption{\label{fig:var}The variance $\sigma_{x}^{2}=\overline{x^{2}}-\overline{x}^{2}$
of the transverse correlation $x=\ln\left|C^{xx}(L/2)\right|$ as
the chain size $L$ is varied for different disorder strengths $D$
for both the uncorrelated and correlated disorder models. The dashed
line is the infinite-randomness prediction that $\sigma_{x}^{2}\sim L$.}
\end{figure}

\begin{figure}[t]
\begin{centering}
\includegraphics[clip,width=0.8\columnwidth]{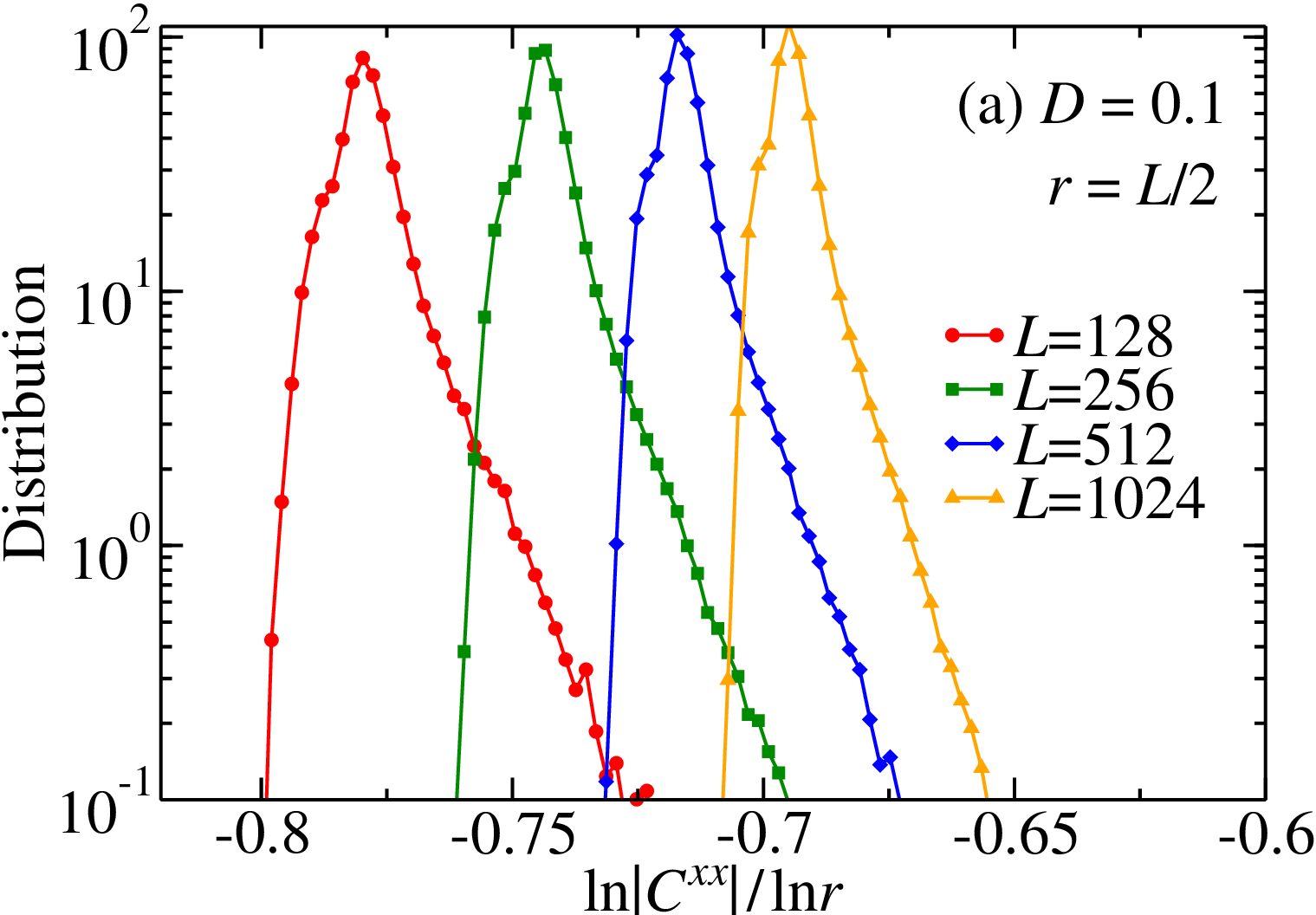}\\
\includegraphics[clip,width=0.8\columnwidth]{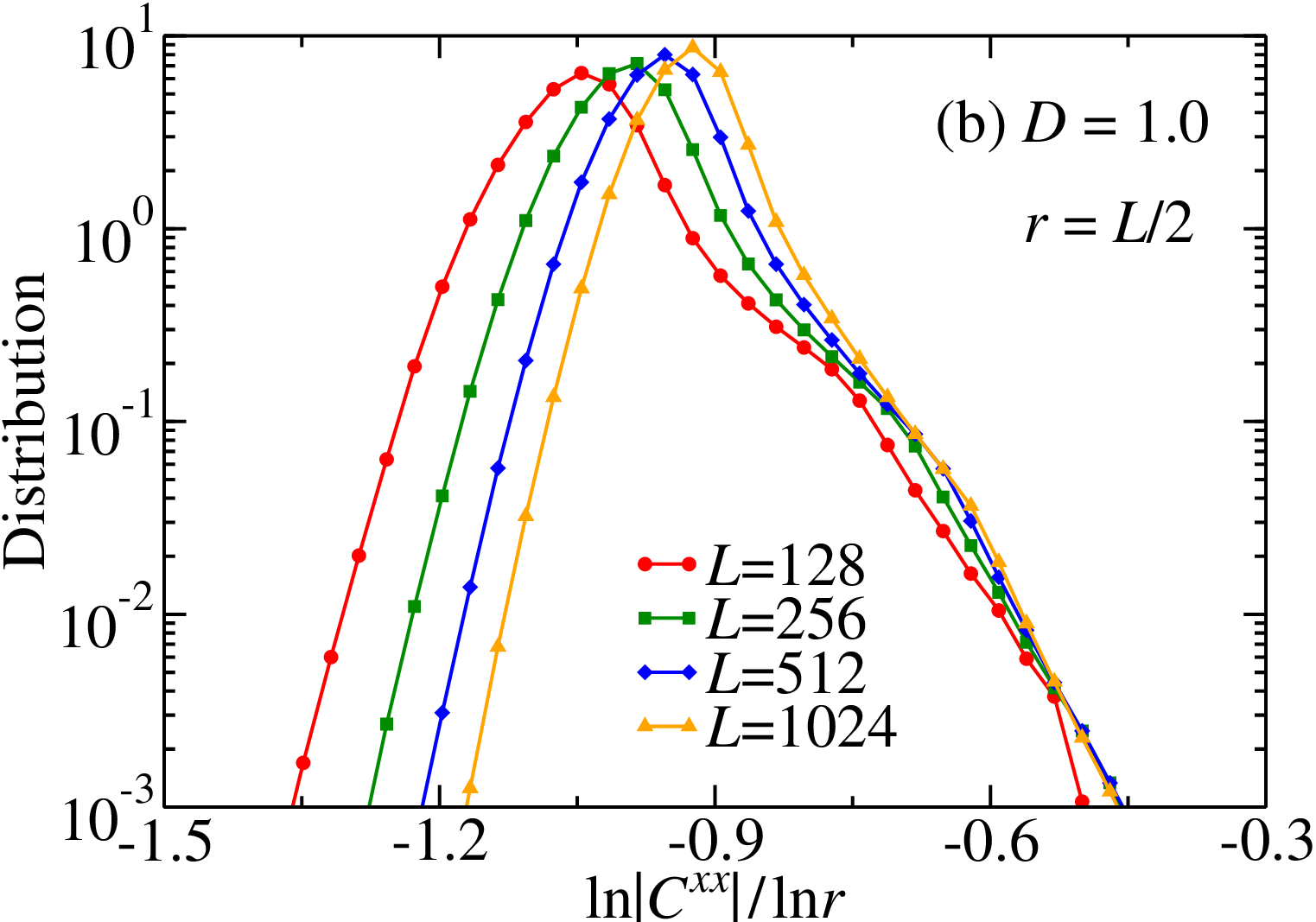} 
\par\end{centering}
\caption{\label{fig:cxdist}Correlated disorder model: The distribution of
the transverse correlation function $C^{xx}(r)$ for disorder strengths
(a) $D=0.1$ and (b) $D=1.0$ and $r=L/2$. The data was obtained
from $N=10^{3}$ disorder realizations for panel (a) and $N=10^{5}$
for panel (b).}
\end{figure}

With respect to the transverse correlation, both typical and average
exponents are equal within our statistical error, and increase monotonically
but is bounded to $1$. This suggests that the distribution of $\ln\left|C^{xx}\right|$
has finite and small width for any distance $r$ and system size $L$
as can be verified in Figs.~\ref{fig:var} and \ref{fig:cxdist}.

\begin{figure}[b]
\begin{centering}
\includegraphics[clip,width=0.8\columnwidth]{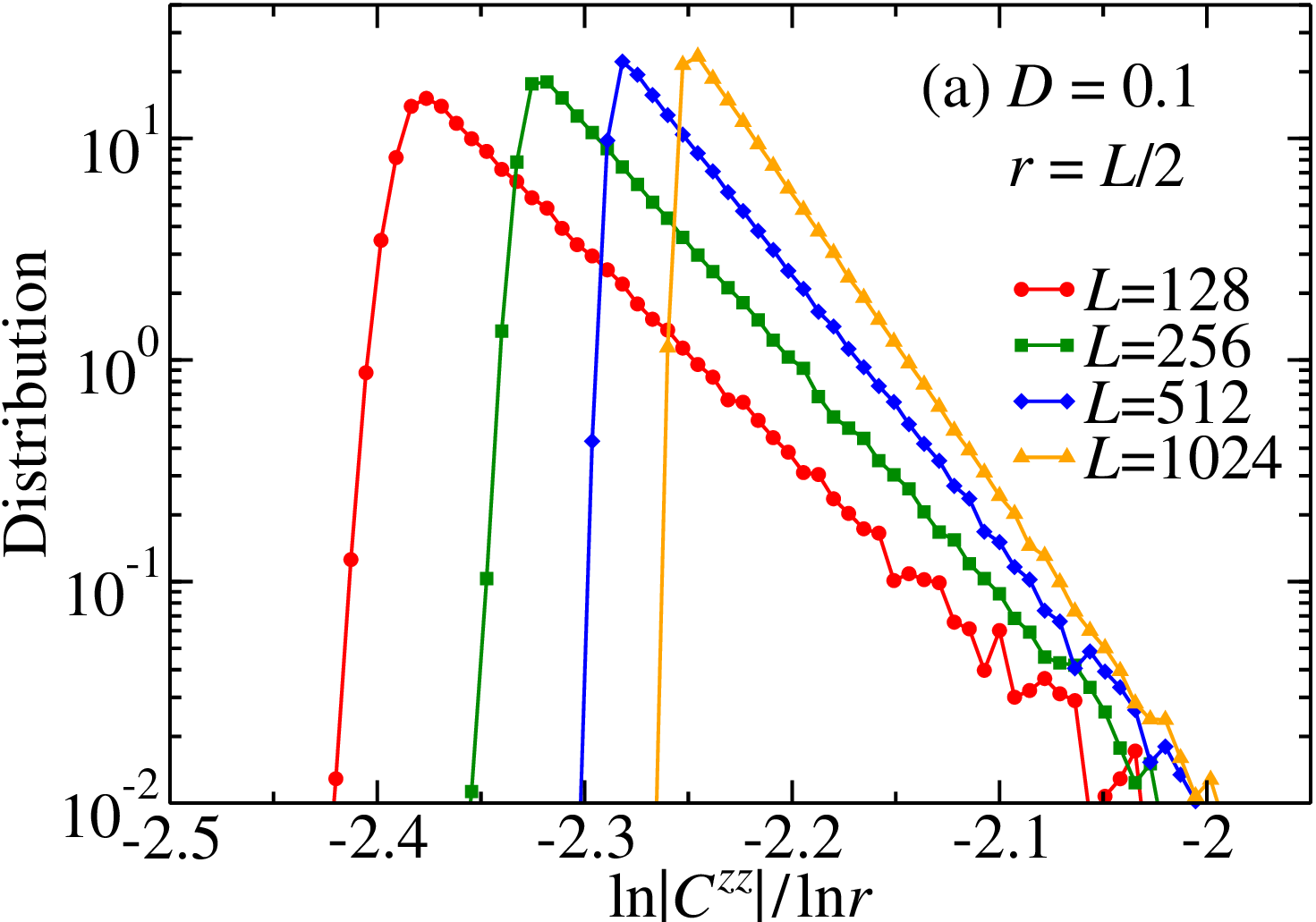}\\
\includegraphics[clip,width=0.8\columnwidth]{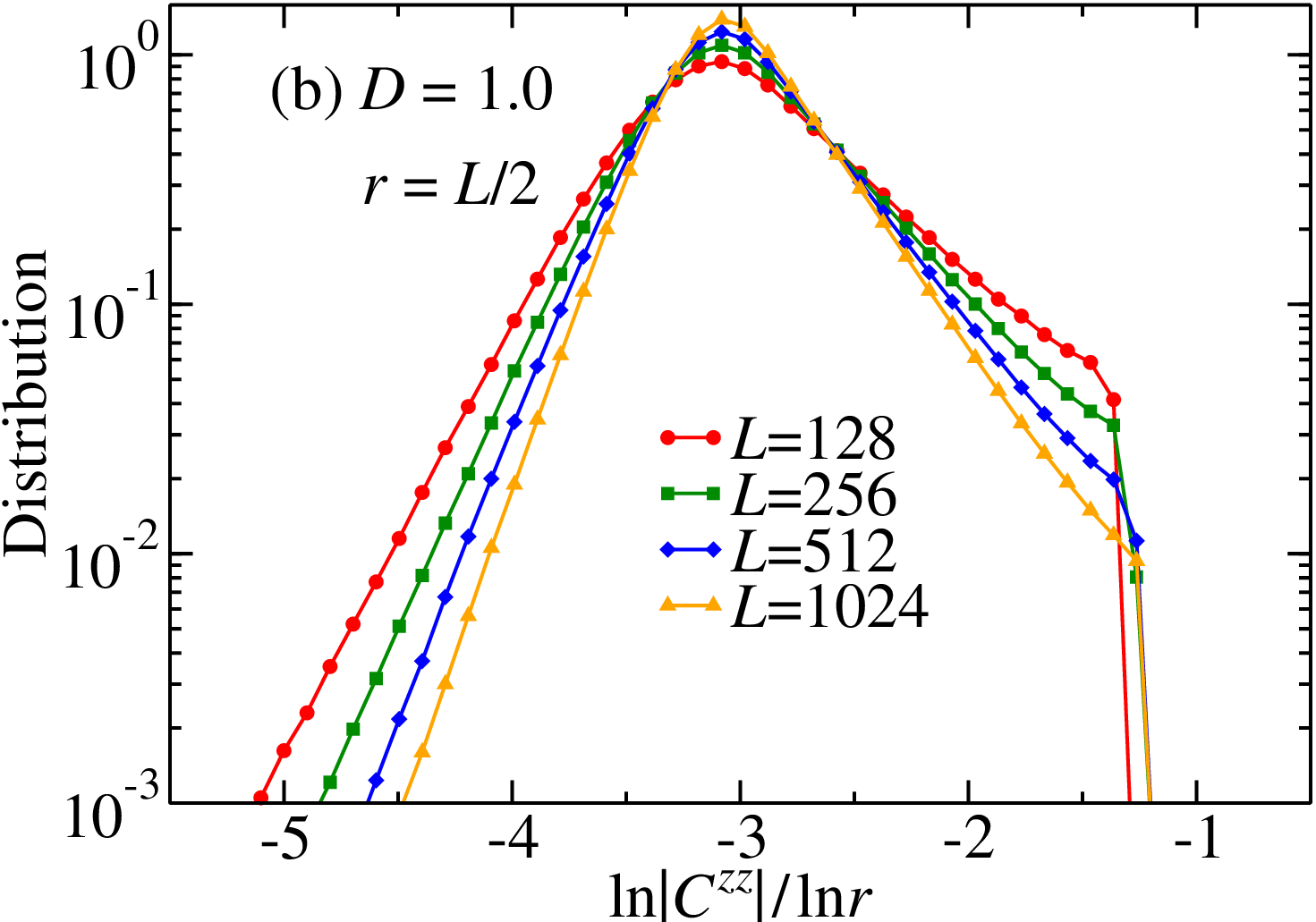}
\par\end{centering}
\caption{\label{fig:czdist}Same as Fig.~\ref{fig:cxdist}, but for the longitudinal
component of the correlation function.}
\end{figure}

In contrast, the typical and average longitudinal correlations behave
quite differently from each other. The average critical exponent remains
equal to the clean one for all disorder strengths studied. The typical
one increases linearly for $D>D_{c}$ apparently without bounds. This
implies that the width of the distribution of $\ln\left|C^{zz}\right|/\ln r$
increases with $D$, as verified in Fig.~\ref{fig:czdist}, but is
fixed for $L$ and $r$ (as we have verified but it is not shown here).

We end this section by calling attention to the striking difference
between transverse and longitudinal correlations. Certainly, the ground
state is far from the random singlet state of the uncorrelated disorder
model. As pointed out in Ref.~\onlinecite{Getelina2016}, the entanglement
properties of the correlated disorder model shares many similarities
with the clean ground state. The fact that typical and average longitudinal
correlations are quite different points towards less similarities.

\section{Conclusions\label{sec:conclusion} }

In this work we have studied the spin-spin correlation functions for
the quantum critical spin-1/2 XX chain in the case of uncorrelated
and correlated coupling constants (see Sec.~\ref{sec:methods}).
In the former case, the chain is governed by a universal (disorder
independent) infinite-randomness fixed point, while in the latter,
it is governed by the clean fixed point for weak disorder ($D<D_{c}$)
and by a line of finite-randomness fixed points tuned by the disorder
strength ($D>D_{c}$) .

For uncorrelated disorder, we have proposed and numerically verified
that the correlations in Eqs.~\eqref{eq:Czz-mean-general}, \eqref{eq:Cxx-general},
\eqref{eq:C-typ-general} are good approximations in the regime $\gamma_{D}r\gg1$
(not restricted to the thermodynamic limit $r\ll L$) for periodic
boundary conditions. We have shown that the chord length \eqref{eq:chord-length}
is not the true scaling variable, exhibiting small corrections for
the mean correlations and even smaller for the typical ones. We have
parameterized and quantified these corrections through the function
$f_{\alpha}$ in Eq.~\eqref{eq:f-correction}. In principle, these
corrections should be non-universal, i.e., disorder dependent. While
this may be indeed the case, we could fit our data using the hypothesis
that $f_{\alpha}$ is universal. Naturally, deciding whether $f_{\alpha}$
is universal or not requires better statistics and larger systems
which are out of our current reach.

In addition, we have studied the corresponding non-universal numerical
prefactors and linked them to the corresponding Lyapunov exponent
\eqref{eq:Lyapunov-exponent} which, ultimately, link them to the
disorder strength. Surprisingly, we have determined an accurate scaling
as quantified in Eqs.~\eqref{eq:prefactor-Cxx2} and \eqref{eq:prefactor-c}
for the mean transverse correlations and the typical longitudinal
and transverse correlations, respectively. In general, these prefactors
and their scaling with a crossover length depend on the dimensions
of the related relevant and irrelevant operators. It is not the scope
of the present work to find those operators and their dimensions.
We leave this as an open question and hope that our findings serve
as future motivation.

We have also studied the distribution of correlations. We have confirmed
(not for the first time) the conjecture that the quantity $\ln\left|C^{\alpha\alpha}\right|/\sqrt{r}$
converges to a non-trivial distribution in the large-separation regime.
In addition, based on the knowledge built from the typical correlation
and its relation with the Lyapunov exponent, we have numerically determined
that, in the large separation limit, the distribution of $\ln\left|C^{\alpha\alpha}\right|/\sqrt{\gamma_{D}r}$
converges to a non-trivial $\alpha$-dependent, narrow and universal
(disorder-independent) distribution quantified in Eq.~\eqref{eq:dist-P}.

It is desirable to generalize our results to other anisotropies $\Delta\neq0$.
It is not entirely clear whether a single-parameter scaling will be
possible for all $-\frac{1}{2}<\Delta\leq1$. Assuming that the SDRG
method is indeed asymptotically exact in this range of anisotropies,
it is then plausible that our results generalize (since $\Delta\rightarrow0$
under the SDRG flow) at the simple cost of correcting the Lyapunov
exponent. Based on the field-theory methods of Ref.~\onlinecite{Giamarchi1989,Doty1992},
it is plausible that Eq.~\eqref{eq:Lyapunov-exponent} generalizes
to $\gamma_{D}\sim D^{\frac{2}{3-2K}}$, with the Luttinger parameter
$K=1-\pi^{-1}\arccos\left(\Delta\right)$. Evidently, the values of
non-universal quantities such as $a_{2n,\alpha}$ in \eqref{eq:f-correction}
may depend on $\Delta$.

Last but not least, we have shown the importance of the finite-size
effect and numerical instabilities when characterizing the random-singlet
state. They are so strong that can mimic logarithmic corrections to
the correct scaling (see Fig.~\ref{fig:sqrt}).

Conversely, for the correlated disorder model we have shown that the
typical correlation functions decay as a power law (see Fig.~\ref{fig:ctyp}),
just like the mean correlations. The corresponding exponents were
determined (see Fig.~\ref{fig:exponent}) and vary continuously for
$D>D_{c}\approx0.3$. While the exponents for the mean and typical
transverse correlations remain nearly equal (implying a narrow distribution
of correlations), the behavior is strikingly different for the longitudinal
correlations, a consequence of the fact that the distribution of longitudinal
correlations are much broader. In addition, we have determined the
chord length is not the correct scaling variable (but it is a very
good approximation to it). The fact that the transverse and longitudinal
correlations behave so differently implies that the random-singlet
state is far from being a good approximation of the true ground state
even when $D\rightarrow\infty.$ The infinite-randomness low-energy
physics of the uncorrelated disorder model is not adiabatically connected
to the strong but finite-randomness behavior of the correlated disorder
model.
\begin{acknowledgments}
We would like to thank Anders Sandvik and Róbert Juhász for useful
discussions, and Nicolas Laflorencie for bringing Ref.~\onlinecite{Shu2016}
to our attention. This study was financed in part by the Coordenação
de Aperfeiçoamento de Pessoal de Nível Superior - Brasil (CAPES) -
Finance Code 001, and also by the Brazilian funding agencies CNPq
and FAPESP.

All authors contributed equally to this work.

\end{acknowledgments}

\bibliographystyle{apsrev4-1}
\bibliography{references}

\end{document}